\documentclass[11pt]{article}

\usepackage{amsthm,amsmath,amssymb,bbm,bm}
\usepackage{natbib}
\usepackage{multirow}
\usepackage[pdftex]{graphicx}

\usepackage{booktabs}
\usepackage{array}
\usepackage{url}
\usepackage{algorithm}
\usepackage{algorithmic}
\usepackage{wrapfig}
\usepackage{lipsum}
\usepackage{mathrsfs}
\usepackage{dsfont}
\usepackage{titling}
\usepackage{caption}

\usepackage[usenames,dvipsnames,svgnames,table]{xcolor}
\usepackage[colorlinks,
linkcolor=red,
anchorcolor=blue,
citecolor=blue
]{hyperref}

\usepackage{geometry}
\usepackage{enumitem}

\newcommand{\source}[1]{\hfill #1} 

\def\T{{ \mathrm{\scriptscriptstyle T} }}

\def\bx{{x}}
\def\bX{{X}}

\newtheorem{step}{Step}
\newtheorem{assumption}{Assumption}
\newtheorem{theorem}{Theorem}
\newtheorem{lemma}{Lemma}

\def\re{{\mathbb R}}

\DeclareMathOperator{\sign}{sign}

\begin{document}

\title{\line(1,0){430}\\{\LARGE Tree based weighted learning for estimating individualized treatment rules with censored data}\\ \line(1,0){430}}

\author{
Yifan Cui\thanks{Department of Statistics and Operations Research, University of North Carolina at Chapel Hill, Chapel Hill, NC  27599, USA; e-mail: \href{mailto:cuiy@live.unc.edu}{cuiy@live.unc.edu}. }
~~Ruoqing Zhu\thanks{Department of Statistics, University of Illinois at Urbana-Champaign, Champaign, IL 61820, USA; email: \href{mailto:rqzhu@illinois.edu}{rqzhu@illinois.edu}.}
~~Michael Kosorok \thanks{Department of Biostatistics and Department of Statistics and Operations Research, University of North Carolina at Chapel Hill, Chapel Hill, NC 27599, USA; email: \href{mailto: kosorok@unc.edu}{kosorok@unc.edu}.}
}

\date{\today}

\maketitle

\begin{abstract}
Estimating individualized treatment rules is a central task for personalized medicine. \cite{zhao2012estimating} and \cite{zhang2012robust} proposed outcome weighted learning to estimate individualized treatment rules directly through maximizing the expected outcome without modeling the response directly. In this paper, we extend the outcome weighted learning to right censored survival data without requiring either an inverse probability of censoring weighting or a semiparametric modeling of the censoring and failure times as done in \cite{zhao2015doubly}. To accomplish this, we take advantage of the tree based approach proposed in \cite{zhu2012recursively} to nonparametrically impute the survival time in two different ways. The first approach replaces the reward of each individual by the expected survival time, while in the second approach only the censored observations are imputed by their conditional expected failure times. We establish consistency and convergence rates for both estimators. In simulation studies, our estimators demonstrate improved performance compared to existing methods. We also illustrate the proposed method on a phase III clinical trial of non-small cell lung cancer.
\end{abstract}

\noindent {\bf keywords}
Individualized treatment rule, Nonparametric estimation, Right censored data, Consistency, Recursively imputed survival trees, Outcome weighted learning.

\section{Introduction}

An individualized treatment regime provides a personalized treatment strategy for each patient in the population based on their individual characteristics. A significant amount of work has been devoted to estimating optimal treatment rules \citep{murphy2003optimal, qian2011performance, zhang2012robust, zhao2011reinforcement, zhao2012estimating}. While each of
these approaches has strengths and weaknesses, we highlight the approach in \cite{zhao2012estimating} because of its robustness to model misspecification (this is similarly true of the approach in \cite{zhang2012robust}) combined with its ability to incorporate support vector machines through the recognition that optimizing the treatment rule can be recast as a weighted classification problem. This approach is commonly referred to as outcome weighted learning. In clinical trials, right censored survival data are frequently observed as primary outcomes. Adapting outcome weighted learning to the censored setting, \cite{zhao2015doubly} proposed two new approaches, inverse censoring weighted outcome weighted learning and doubly robust outcome weighted learning, both of which require semiparametric estimation of the conditional censoring probability given the patient characteristics and treatment choice. The doubly robust estimator additionally involves semiparametric estimation of the conditional failure time expectation but only requires that one of the two models, for either the failure time or censoring time, be correct. Potential drawbacks of these methods are that either or both models may be misspecified and inverse censoring weighting estimation can be unstable numerically \citep{qian2011performance, zhu2012recursively}.

In this paper, we propose a nonparametric tree based approach for right censored outcome weighted learning which avoids both the inverse probability of censoring weighting and restrictive modeling assumptions for imputation through recursively imputed survival trees \citep{zhu2012recursively}. Since the true failure times $T$ are only partially known, they cannot be used directly as weights in the outcome weighted learning \citep{zhao2012estimating} framework. However, recursively imputed survival trees \citep{zhu2012recursively} provide an alternative approach to weighting by using the conditional expectations of censored observations without requiring inverse weighting. Tree-based methods \citep{breiman1984classification, breiman2001random} are a broad class of nonparametric estimators which have become some of the most popular machine learning tools. Its adaptation to the survival setting has also drawn a lot of interests in the literature \citep{leblanc1992relative, hothorn2004bagging, ishwaran2008random}, and it has also been used for interpretable prediction modeling in personalized medicine \citep{laber2015tree}. The recursively imputed survival tree approach \citep{zhu2012recursively} combines extremely randomized trees with a recursive imputation method, which has been shown to improve performance and reduce prediction error while avoiding estimation of inverse censoring weights without making parametric or semiparametric assumptions on the conditional probability distribution of the failure time. Numerical studies demonstrate that the proposed method outperforms existing alternatives in a variety of settings.

The proposed method uses these recursively imputed survival trees to impute the survival times nonparametrically in a manner suitable for implementation within outcome weighted learning. We verify this novel approach both theoretically and in numerical examples. As part of this, we also present for the first time consistency and rate results for tree-based survival models in a more general setting than the categorical predictors considered in \cite{ishwaran2010consistency}.

The remainder of the article is organized as follows. In section \ref{section2}, we present the mathematical framework for individualized treatment rules for right censored survival outcomes. In section \ref{section3} we establish consistency and an excess value bound for the estimated treatment rules. Extensive simulation studies are presented in Section \ref{section4}. We also illustrate our method using a phase III clinical trial on non-small cell lung cancer in Section \ref{section5}. The article concludes with a discussion of future work in Section \ref{section6}. Some needed technical results are provided in the Appendix.

\section{Methodology}\label{section2}

\subsection{Individualized treatment regime framework}

Before characterizing the individualized treatment regime, we first introduce some general notation and introduce the value function, and then extend the notation and ideas to the censored data setting. Let $X \in \cal {X}$ be the observed patient-level covariate vector, where $\mathcal{X}$ is a $d$ dimensional vector space, and let $A\in \{-1, +1\}$ be the binary treatment indicator. $\widetilde{T}$ is the true survival time, however, we consider a truncated version at $\tau$, i.e., $T=\min(\widetilde{T},\tau)$, where the maximum follow-up time $\tau<\infty$ is a common practical restriction in clinical studies. The goal in this framework is to maximize a reward $R$, which could represent any clinical outcome. Specifically, we wish to identify a treatment rule $\cal{D}$, which is a map from the patient-level covariate space $\mathcal{X} $ to the treatment space $\{+1,-1\}$ which maximizes the expected reward.
In the survival outcome setting, we use $R=T$ or $\log(T)$ as done in \cite{zhao2015doubly}.

To achieve this maximization, we define the value function as
$$V(\mathcal{D}) = E^{\mathcal{D}}(R) = E \big[RI\{A={\mathcal D}(X)\}/\pi(A;X) \big],$$
where $I\{\cdot\}$ is an indicator function, $\pi(a;X)=\text{pr}(A=a \mid X)>M'~ a.s.$ for some $M'>0$ and each $a\in \{+1,-1\}$. The function $\pi$ is the propensity score and is known in a randomized trial setting, which we assume is the case for this paper, but needs to be estimated in a non-randomized, observational study setting. The individualized treatment regime we are most interested in is the optimal treatment rule $\cal{D}^*$ which maximizes the value function, i.e.
\begin{align}
\mathcal{D}^* = \underset{\cal{D}}{\arg\max} \,\, E\big[ RI\{A={\mathcal D}(X)\}/\pi(A;X) \big]. \label{owl1}
\end{align}
After rewriting the value function as
\begin{align*}
V(\mathcal{D})=E\big[ E(R \mid A=1,X)I\{\mathcal{D}(X)=1\}+E(R \mid A=-1,X)I\{\mathcal{D}(X)=-1\} \big],
\end{align*}
it is easy to see that
\begin{align*}
\mathcal{D}^*=  \sign \big\{ E(R \mid A=1,X)-E(R \mid A=-1,X) \big\}.
\end{align*}
Hence, the definition of $\mathcal{D}^*$ is equivalent to $\mathcal{D}^*(x) = \arg\max_{a} E(R \mid A=a,X=x)$. Instead of maximization the objective function in \eqref{owl1}, the outcome weighted learning approach searches for the optimal decision rule $\mathcal{D}^*$ by minimizing the weighted misclassification error, i.e.,
\begin{align}
\mathcal{D}^* = \underset{\cal{D}}{\arg\min} \,\, E\big[ RI\{A \neq {\mathcal D}(X)\}/\pi(A;X) \big]. \label{owl2}
\end{align}
In an ideal situation, we would replace $R$ with $T$ or $\log(T)$. However, this is not possible under right censoring.

\subsection{Value function under right censoring}

Consider a censoring time $C$ that is independent of $T$ given $(X, A)$. We then have the observed time $Y = \min(T, C)$, and the censoring indicator $\delta= I(T \leq C)$. Assume that $n$ independent and identically distributed copies, $\{Y_i, \delta_i, X_i, A_i\}_{i=1}^{n}$, are collected. Since $T$ is not fully observed we seek for a sensible replacement which maintains as close as possible the same value function. We propose two approaches in the following, denoted as $R_1$ and $R_2$ respectively. The first approach is to obtain a nonparametric estimated conditional expectation $\widehat E(T \mid X, A)$. Letting $R_1=E(T \mid X,A)$ and bringing the expectation of $T$ inside, we have
\begin{align}
E \big[T I\{A={\mathcal D}(X)\}/\pi(A;X) \big] = E \big[ R_1 I\{A={\mathcal D}(X)\}/\pi(A;X) \big]. \label{value_r1}
\end{align}
Another approach is to replace only the censored observations conditioning on the observed data. It is interesting to observe that the conditional expectation of $T$, given $Y$ and $\delta$, can be written as
\begin{align}
R_2 :=& E(T \mid X, A, Y, \delta) \nonumber \\
=& I(\delta = 1) Y + I(\delta = 0) E(T \mid X, A, Y, \delta = 0) \nonumber \\
=& I(\delta = 1) Y + I(\delta = 0) E(T \mid X, A, C = Y, T > Y, Y) \nonumber \\
=& I(\delta = 1) Y + I(\delta = 0) E(T \mid X, A, T > Y, Y).  \label{value_r2}
\end{align}
An important property that we used in the last equality is the conditional independence between $T$ and $C$. With the information of $Y=y$ given, and knowing that $\delta = 0$, the conditional distribution of $T$ is defined on $(c, \tau]$ with density function proportional to the original density of $T$. In other words, the conditional survival function of $T$ is $S(t \mid X, A)/S(c \mid X, A)$ for $t > c$, where $S(\cdot \mid X, A)$ is the conditional survival function of $T$. Hence, we can calculate the expectation of $T$ accordingly. With the definition of $R_2$, it is easy to see that the corresponding value function is equivalent to the left side of equation (\ref{value_r1}) by further taking expectations with respect to $Y$ and $\delta$. Note that the above arguments remain unchanged if we replace $T$, $C$ and $Y$ with $\log(T)$, $\log(C)$, and $\log(Y)$, respectively: this equivalence will be tacitly utilized throughout the paper, except when the distinction is needed.

With our proposed two reward measures, the remaining challenge is to nonparametrically estimate the conditional expectations. To this end, we utilize the nonparametric tree based method proposed by \cite{zhu2012recursively}. It is worth noting that the conditional expectation of $T$ defined in $R_2$ shares the same logical underpinnings as the imputation step in \cite{zhu2012recursively}. However, the goal of the imputation step is to replace the censored observations with a randomly generated conditional failure time which utilizes the same condition survival distribution of $T$ given $T > C$. We will provide details of the estimation procedure in the next section. To conclude this section, we provide the empirical versions of the value function using the two rewards $R_1$ and $R_2$, respectively, which we solve for the optimal decision $\cal D^*$ by minimization:
\begin{align}
&  n^{-1}\sum_{i=1}^{n} \frac{\widehat E(T_i \mid A_i,X_i) I\{A_i={\mathcal D}(X_i)\}}{\pi(A_i;X_i)}, \label{hard1} \\
\text{and \,\,} n^{-1}\sum_{i=1}^{n} & \frac{\{ \delta_i Y_i+(1-\delta_i) \widehat E(T_i \mid X_i, A_i, T_i > Y_i, Y_i )\} I\{A_i={\mathcal D}(X_i)\}}{\pi(A_i;X_i)}. \label{hard2}
\end{align}
\subsection{Outcome weighted learning with survival trees}

The recursively imputed survival trees method proposed by \cite{zhu2012recursively} is a powerful tool to estimate conditional survival functions for censored data. A brief outline of the algorithm is provided in the following. We refer interested readers to the original paper for details. To fit the model, we first generate extremely randomized survival trees for the training dataset. Secondly, we calculate conditional survival functions for each censored observation, which can be used for imputing the censored value to a random conditional failure time. Thirdly, we generate multiple copies of the imputed dataset, and one survival tree is fitted for each dataset. We repeat the last two steps recursively and the final nonparametric estimate of $\widehat E(T \mid X,A)$ is obtained by averaging the trees from the last step.

Following \citep{zhao2012estimating}, we next use support vector machines to solve for the optimal treatment rule. A decision function $f(x)$ is learned by replacing $I\{A_i={\mathcal D}(X_i)\}$ in Equations \eqref{hard1} or \eqref{hard2} with $\phi\{A_i f(X_i)\}$, where $\phi(x)=(1-x)^+$ is the hinge loss and $x^+=\max(x,0)$. Furthermore, to avoid overfitting, a regularization term $\lambda_n \|f\|^2$ is added to penalize the complexity of the estimated decision function $f$. Here, $\| f\|$ is some norm of $f$ , and $\lambda_n$ is a tuning parameter. A high-level description of the proposed method is given in Algorithm \ref{algo1} below. We consider both linear and nonlinear decision functions $f$ when solving (\ref{learningopt}). For a linear decision function, $f(\bx)=\theta_0 + \theta^T \bx$ and we let $\|f\|$ be the Euclidean norm of $\theta$. For nonlinear decision functions, we employ a universal kernel function $k : \mathcal{X} \times\mathcal{X} \rightarrow \re$, such as the Gaussian kernel, which is continuous, symmetric and positive semidefinite. The optimization problem is then equivalent to a dual problem that maximizes
\begin{align*}
\sum_{i=1}^n \alpha_i - \frac{1}{2} \sum_{i=1}^n \sum_{j=1}^n \alpha_i\alpha_jA_iA_jk(\bX_i,\bX_j),
\end{align*}
subject to $0\leq\alpha_i\leq \gamma W_i/\pi_i$ and $\sum_{i=1}^n\alpha_iA_i=0$, where $W_i$ is the numerator in either (\ref{hard1}) or (\ref{hard2}) and $\pi_i$ is the respective denominator. Both settings can be efficiently solved by quadratic programming. For further details regarding solving weighted classification problems using support vector machines, we refer to \citep{zhao2012estimating,zhao2015doubly,chang2011libsvm}.

\begin{algorithm}  \caption{Pseudo algorithm for the proposed method} \label{algo1}
\end{algorithm}
\begin{step}
Use $\{(X_i^\T, A_i, A_iX_i^\T)^\T, Y_i, \delta_i\}_{i=1}^n$ to fit recursively imputed survival trees. Obtain the estimation $\widehat E(T_i \mid A_i, X_i)$ for reward $R_1$ or the estimation $\widehat E(T_i \mid X_i, A_i, T_i > Y_i, Y_i)$ for reward $R_2$.
\end{step}
\begin{step}
Let the weights $W_i$ be either $\widehat E(T_i \mid A_i, X_i)$ or $\delta_i Y_i+(1-\delta_i)\widehat E(T_i \mid A_i,X_i,T_i > Y_i, Y_i)$, depending on which of the two proposed approaches is used. Minimize the following weighted misclassification error:
\begin{equation}
\widehat f(x) = \underset{f}{\arg\min} \sum_{i=1}^n W_i \frac{\phi\{A_i f(X_i)\}}{{\pi(A_i;X_i)}}+\lambda_n \| f\|^2.
\label{learningopt}
\end{equation}
\end{step}
\begin{step}
Output the estimated optimal treatment rule $\widehat {\cal D}(x)=\textrm{sign}\{\widehat f(x)\}.$
\end{step}

\section{Theoretical results}\label{section3}
\subsection{Preliminaries}
The risk function is defined as
\begin{align*}
R(f)=E\Big[ \frac{R}{\pi(A;X)}I\{A \neq \sign(f(X))\} \Big],
\end{align*}
where the reward $R = R_1 = E(T \mid X, A)$ for the first approach, or $R = R_2 = \delta Y + (1-\delta)E(T \mid X,A, T > Y, Y)$ for the second one. We define $\phi$-risk for both the true and the working model as, respectively, $R_{\phi}(f)=E[R\phi\{Af(X)\}/\pi(A;X)]$ and $R'_{\phi}(f)=E[\widehat R\phi\{Af(X)\}/\pi(A;X)]$, where $\hat{R}$ is the estimated value of $R$ based on one of the two proposed methods. We also define the hinge loss function for the true and working models as $L_\phi(f)=R\phi\{Af(X)\}/\pi(A;X)$ and $L'_\phi(f)=\widehat R\phi\{Af(X)\}/\pi(A;X)$, respectively.

The proposed estimator $\widehat{\mathcal{D}}=\sign(\widehat{f}_n(X))$, where $\widehat f_n$ is solved by one of the following optimization problems within some reproducible kernel Hilbert space $\mathcal{H}_k$:
\begin{align*}
\hat{f}_n &= \underset{f\in \mathcal{H}_k}{\arg\min} \,\, n^{-1} \sum_{i=1}^n \frac{\widehat E(T_i \mid X_i,A_i)}{\pi(A_i;X_i)} \phi\{f(X_i)A_i\}+\lambda_n ||f||_n^2,
\end{align*}
or
{\small
\begin{align*}
\hat{f}_n &= \underset{f\in \mathcal{H}_k}{\arg\min} \,\, n^{-1} \sum_{i=1}^n \frac{ \delta_i Y_i+ (1-\delta_i)\widehat E(T_i \mid X_i,A_i, T_i>Y_i, Y_i)}{\pi(A_i;X_i)} \phi\{f(X_i)A_i\}+\lambda_n ||f||_n^2.
\end{align*}
}

\subsection{Consistency of tree-based survival models}\label{proof2}

In this section, we provide the convergence bound of a simplified tree-based survival model, which is very close to the original algorithm in \cite{zhu2012recursively}. The purpose of this section and its main result, Theorem \ref{tree}, is to demonstrate the existence of an accurate estimator of the underlying hazard function when tree-based methods are used. An earlier result developed in \cite{ishwaran2010consistency} considers only categorical feature variables. To the best of our knowledge, what we present below is the first consistency result for a tree-based survival model under general settings with restrictions only on the splitting rules, which is interesting in its own right.

For simplicity, we assume in this section that $\mathcal{Q}_n=\{(Y_i,\delta_i,X_i,A_i), i=1,\ldots,n\}$ is the training sample, where $X_i$ is independent uniformly distributed on $[0,1]^d$. The result can be easily generated to distributions with bounded support and density function bounded above and below. For any fixed $X$, our goal is to estimate the cumulative hazard function of failure time $r(\cdot, X,A)=\Lambda_T(\cdot \mid X, A)$; hereinafter, we write it as $\Lambda(\cdot \mid X, A)$.

A random forest is a collection of randomized regression trees $\{\hat r_n(\cdot, X, A,\\ \Theta_j,\mathcal{Q}_n), 1\leq j\leq m\}$, where $m$ is the number of trees. The randomizing variable $\Theta$ is used to indicate how the successive cuts are performed when an individual tree is built. Hence the forest version of the survival tree model can be expressed as
\begin{align*}
\hat r_n(\cdot, X, A, \mathcal{Q}_n) = \frac{1}{m}\sum_{j=1}^m \hat r_n(\cdot, X, A, \Theta_j, \mathcal{Q}_n).
\end{align*}

Here, we consider a simplified scenario in which the selection of the coordinate is completely random and independent from the training data \citep{biau2012analysis}. We only consider the consistency of a single tree and denote our tree estimator as $\hat r_n(\cdot,X,A)$. The result can be easily extended to the situation where $m$ is finite.

A brief description of how each individual tree is constructed is provided in the appendix. Here we highlight some key assumptions and the main result. Our first assumption puts a lower bound on the probability of observing a failure at $\tau$, and the second one assumes the smoothness of the hazard and cumulative hazard functions.

\begin{assumption}\label{addas}
For some $M>0$, $S_Y(\tau \mid X,A)>M$ almost surely.
\end{assumption}

\begin{assumption}\label{LC}
For any fixed time point $t$ and treatment decision $A$, the cumulative hazard function $\Lambda(t\mid X,A)$ is $L$-Lipschitz continuous in terms of $X$, and the hazard function $\lambda(t\mid X,A)$ is $L'$-Lipschitz continuous in terms of $X$, i.e., $|\Lambda(t \mid X_1,A)-\Lambda(t \mid X_2,A)| \leq  L||X_1-X_2||$ and $|\lambda(t \mid X_1,A)-\lambda(t \mid X_2,A)|\leq  L'||X_1-X_2||$, respectively, where $||\cdot||$ is the Euclidean norm.
\end{assumption}

The following theorem provides the bound of the proposed tree based survival model for each $X$. Details of the proof are collected in the Appendix.

\begin{theorem}\label{tree}
Assume that Assumptions \ref{addas}--\ref{LC} and the construction of a tree-based survival model described in the Appendix. Further assume that $k_n\rightarrow \infty$ and $n/k_n\rightarrow \infty$ as $n \rightarrow \infty$, where $k_n$ is a deterministic parameter which we can control  (each individual tree has approximately $k_n$ terminal nodes). We have for each $X$,
\begin{align*}
\text{pr}\Big\{\sup_{t<\tau} |\hat r_n(t, \bX,A)-r(t, \bX,A)| &\leq C [d^{1/2}2^{-\{(1-r)\left \lceil{\log_2 k_n}\right \rceil \}/d}\\
&+b^{1/2}\{(1-u)n 2^{-\lceil \log_2 k_n \rceil}\}^{-1/2}]\Big\} \geq 1 - w_n,
\end{align*}
where $r, u \in (0, 1)$, $b>1/228$, $(1-u)n 2^{-\lceil \log_2 k_n \rceil}\geq 288b/M^4$, $C$ is some universal constant and
\begin{align*}
w_n=~ 16[(1-u)n 2^{-\lceil \log_2 k_n \rceil}+2]e^{-b} + e^{-u^2 n 2^{-\lceil \log_2 k_n \rceil-1}} + de^{-\left \lceil{\log_2 k_n}\right \rceil  r^2/(2d)}.
\end{align*}
\end{theorem}

The ideal balance happens when $k_n=n^{d/(d+2)}$. In this case, the optimal rate of the bound is close to $n^{-1/(d+2)}$. The following theorem proves consistency of the proposed tree based survival model. Details of the proof are collected in the Appendix.

\begin{theorem}\label{tree2}
Assume that Assumptions \ref{addas}--\ref{LC} and the construction of a tree-based survival model described in the Appendix. Further assume that 
$k_n=n^{\eta}$, where $0<\eta<1$. Then the estimator of the survival tree model is consistent. Moreover,
\begin{align*}
\sup_{t<\tau} &E_X |\hat r_n(t, \bX,A)- r(t, \bX,A)| \leq C [d^{1/2}2^{-\{(1-r)\left \lceil{\log_2 k_n}\right \rceil \}/d}\\
&+b^{1/2}\{(1-u)n 2^{-\lceil \log_2 k_n \rceil}\}^{-1/2}+w_n \ln(n)],
\end{align*}
where $r, u \in (0, 1)$,  $b>1/228$, $(1-u)n 2^{-\lceil \log_2 k_n \rceil}\geq 288b/M^4$, $C$ is some universal constant and
\begin{align*}
w_n=~ 16[(1-u)n 2^{-\lceil \log_2 k_n \rceil}+2]e^{-b} + e^{-u^2 n 2^{-\lceil \log_2 k_n \rceil-1}} + de^{-\left \lceil{\log_2 k_n}\right \rceil  r^2/(2d)}.
\end{align*}
\end{theorem}

\subsection{Consistency and Excess Value Bound}
Fisher consistency follows directly from Proposition 3.1 in \cite{zhao2012estimating}, hence the proof is omitted. Here we restate the result as the following lemma. For the proposed method, we simply replace the reward $R$ in $R_{\phi}(f)$ with $R_1$ or $R_2$. Note that both versions are equivalent to the reward function $R_{\phi}(f) = E[T\phi\{Af(X)\}/\pi(A;X)]$:

\begin{lemma}[Proposition 3.1 in \cite{zhao2012estimating}] \label{lemma0}
For any measurable function $\widetilde f$ , if $\widetilde{f}$ minimizes $R_{\phi}(f)$, then $\mathcal{D}^*(x) =\sign(\widetilde{f}(x))$.
\end{lemma}

Provided the Assumptions in Section \ref{proof2} hold, the following lemma ensures the convergence of the estimated conditional expectations. The proof is given in Appendix.

\begin{lemma}\label{lemma1}
Based on Theorem \ref{tree}, for each $X$ the estimated conditional expectations converge in probability, i.e.,
 \begin{align*}
\text{pr} \Big\{ \big|\widehat E(T &\mid X,A) - E(T \mid X,A) \big| \\
&\leq C_1[2^{-\{(1-r)\left \lceil{\log_2 k_n}\right \rceil \}/d}+(b/\{(1-u)n 2^{-\lceil \log_2 k_n \rceil}\})^{1/2}] \Big\}\geq 1-w_n,
\end{align*}
 \begin{align*}
\text{pr} \Big\{ \big| \widehat E(T &\mid X,A,T>Y,Y) - E(T  \mid X,A,T>Y,Y) \big| \\
\leq &C_2[2^{-\{(1-r)\left \lceil{\log_2 k_n}\right \rceil \}/d} +(b/\{(1-u)n 2^{-\lceil \log_2 k_n \rceil}\})^{1/2}]\Big\} \geq 1-2w_n,
\end{align*}
for some constant $C_1$, $C_2$ (depending on $L,L',\tau,M,d$).
\end{lemma}

We will use the above lemmas to prove our main theorem based on the Gaussian kernel. Before we derive the convergence rate and excess value bound, we define the value function corresponding to the true and working model as $V(f) = E( RI[A = \sign\{f(X)\}]/\pi(A;X))$, $V'(f) = E(\widehat RI[A = \sign\{f(X)\}]/\pi(A;X))$, respectively. We further define the empirical $L_2$--norm, $\|f-g\|_{L_2(P_n)}=( \sum_{i=1}^n\\  |f(X_i-g(X_i))|^2/n)^{1/2}$, which also defines an $\epsilon$-ball based on this norm. By Theorem 2.1 in \cite{steinwart2007fast}, we restate the bound for covering numbers:
\begin{lemma}[Theorem 2.1 in \cite{steinwart2007fast}] \label{as3}
For any $\beta>0$, $0< v<2$, $\varepsilon>0$ we have $\sup_{P_n} \log N(B_{\mathcal{H}_k},\varepsilon, L_2(P_n)) \leq c_{v,\beta,d}\sigma_n^{(1-v/2)(1+\beta)d} \varepsilon^{-v}$, where $B_{\mathcal{H}_k}$ is the closed unit ball of $\mathcal{H}_k$, and $d$ is the dimension of $\mathcal{X}$.
\end{lemma}

Lastly, for $\widetilde f=\arg\min_{f\in \mathcal{F}} E\{L_{\phi}(f)\}$, we define the approximation error function
\begin{align*}
a(\lambda)=\inf_{f\in \mathcal{H}_k} [ E\{L_{\phi}(f)\}+\lambda ||f||_k^2 -E\{L_{\phi}(\widetilde f)\}].
\end{align*}

Then we have following theorem, the proof of which is given in Appendix.
\begin{theorem}\label{thm3}
Based on Theorem \ref{tree2} and assuming that the sequence $\lambda_n > 0$ satisfies $\lambda_n \rightarrow 0$ and $\lambda_n \ln n \rightarrow\infty$, we have that
\begin{align*}
\text{pr}(V(f^*) \leq V(\widehat{f}_n) +\epsilon) &\geq 1-2e^{-\rho},
\end{align*}
where $f^*$ maximize the true value function $V$, $\epsilon =~a(\lambda_n)+M_v (n\lambda_n/c_n)^{-2/(v+2)}+M_v\lambda_n^{-1/2}(c_n/n)^{2/(d+2)}+K \rho(n\lambda_n)^{-1}+2K\rho n^{-1}\lambda_n^{-1/2}+C\lambda_n^{-1/2}\{2^{-(1-r)\left \lceil{\log_2 k_n}\right \rceil /d}\\+(b/\{(1-u)n 2^{-\lceil \log_2 k_n \rceil}\})^{1/2}+w_n \ln n\}$, $c_n=c_{v,\beta,d}\sigma_n^{(1-v/2)(1+\beta)d}$ and $\rho>0$ for both methods; also, $M_v$ is a constant depending on $v$, $K$ is a sufficiently large positive constant and $C$ is a some large constant depending on $d$.
\end{theorem}

The rate consists of two parts. The first part is from the approximation error using $\mathcal{H}_k$. The second part controls the approximation error due to using the proposed tree-based method to estimate the conditional expectation.

\section{Simulation studies}\label{section4}
We perform simulation studies to compare the proposed method with existing alternatives, including the Cox proportional hazards model with covariate-treatment interactions, inverse censoring weighted outcome weighted learning, and doubly robust learning, both proposed in \citep{zhao2015doubly}. We use survival time on the log scale $\log(T)$ as outcome. We also present for comparison an ``oracle'' approach which uses the true failure time on the log scale $\log(T)$ as the weight in outcome weighted learning, although this would not be implementable in practice. However, this approach is a representation of the best possible performance under the outcome weighted learning framework.

We generate $X_i$'s independently from a uniform distribution. Treatments are generated from $\{+1,-1\}$ with equal probabilities. We present four scenarios in this simulation study. The failure time $T$ and censoring time $C$ are generated differently in each scenario, including both linear and nonlinear decision rules. For each case, we learn the optimal treatment rule from a training dataset with sample size $n=200$. A testing dataset with size 10000 is used to calculate the value function under the estimated rule. Each simulation is repeated 500 times.

Tuning parameters in the tree based methods need to be selected. We mostly use the default values. The number of variables considered at each split is the integer part of the square root of $d$ as suggested by \cite{ishwaran2008random} and \cite{geurts2006extremely}. We set the total number of trees to be 50 as suggested by \cite{zhu2012recursively} and use one fold imputation. For the alternative approaches such as inverse censoring weighted outcome weighted learning and doubly robust learning, a Cox proportional hazards model with covariates $(X, A, XA)$ is used to model $T$ and $C$ respectively. Note that when at least one of the two working models is correctly specified, the doubly robust method enjoys consistency. We implemented outcome weighted learning using a Matlab library for support vector machine \citep{chang2011libsvm}. Both linear and Gaussian kernels are considered for all methods except for the Cox model approach which could be directly inverted to obtain the decision rules. The parameter $\lambda_n$ is chosen by ten-fold cross-validation.

\subsection{Simulation settings}

For all scenarios, we generate $\widetilde{T}$ and $C$ independently. The failure time $T=\min(\tau,\widetilde T)$. For all accelerated failure time models, $\epsilon$ is generated from a standard normal distribution. For all Cox proportional hazards models, the baseline hazard function $\lambda_0(t) = 2t$. For all simulation results presented in this section, we consider setting the censoring rates to approximately 45\% for all scenarios. We also perform a sensitivity analysis for different censoring rates (30\% and 60\%) for each scenario. These additional results are presented in the Appendix.

Scenario 1. Both $\widetilde{T}$ and $C$ are generated from the accelerated failure time model. $\tau=2.5$ and $d = 10$. The optimal decision function is linear. The value of the optimal treatment rule is approximately 0.031:

\begin{align*}
\log(\widetilde{T})=&-0.2-0.5X_1+0.5X_2+0.3X_3 \nonumber \\
&+(0.5-0.1X_1-0.6X_2+0.1X_3)A+\epsilon, \\
\log(C)=&0.1-0.8X_1+0.4X_2+0.4X_3+(0.5-0.1X_1-0.6X_2+0.3X_3)A+\epsilon.
\end{align*}

Scenario 2. $\widetilde T$ is generated from a Cox model and $C$ is generated from the accelerated failure time model. The optimal decision function is nonlinear. $\tau=8$ and $d=10$. The value of the optimal treatment rule is approximately 0.181:
\begin{align}
\lambda_{\widetilde{T}}(t \mid A,X)=&\lambda_0(t)\exp\{-0.2-1.5X_1^{1.5}+0.5X_2+(0.8-0.7X_1^{0.5}-1.2X_2^{2})A\}, \nonumber \\
\log(C)=&-0.5+0.7X_1+X_2^2+0.6X_3+0.1X_4 \nonumber \\
&+ (0.2+X_1^{2.5}-2X_2+0.5X_3)A + \epsilon. \nonumber
\end{align}

Scenario 3. $\widetilde T$ is generated from an accelerated failure time model with tree structured effects. $C$ is generated from a Cox model with nonlinear effects. $\tau=8$ and $d=5$. The value of the optimal treatment rule is approximately 1.079:
\begin{align}
\log(\widetilde{T})=&X_1+I(X_2>0.5)I(X_3>0.5)+ (0.3-X_1) A \nonumber \\
&+2\{I(X_4<0.3)I(X_5<0.3)+I(X_4>0.7)I(X_5>0.7)\}A+\epsilon, \nonumber \\
\lambda_{C}(t \mid A,X)=&\lambda_0(t)\exp\{-1.5+X_1+(1+0.6X_2^{1.5})A\}. \nonumber
\end{align}


Scenario 4. $\widetilde T$ is is generated from an accelerated failure time model. $C$ is generated from a Cox model. $\tau=2$ and $d=10$. The value of the optimal treatment rule is approximately -0.389:
\begin{align}
\log({\widetilde T}) =& -0.5 - 0.8X_1+0.7X_2+0.2X_3 \nonumber \\
&+ (0.6-0.4X_1-0.2X_2-0.4X_3)A + \epsilon, \nonumber \\
\lambda_C(t \mid A,X) =& \lambda_0(t)\exp\{-0.5 X_1-0.5X_2+0.2X_3 \nonumber \\
&-(1-0.5X_1+0.3X_2-0.5X_3)A\}.\nonumber
\end{align}

\subsection{Simulation results}

\begin{figure}[H]
\centering
   \includegraphics[trim=0in 0in 0in 0in, height=1.6in]{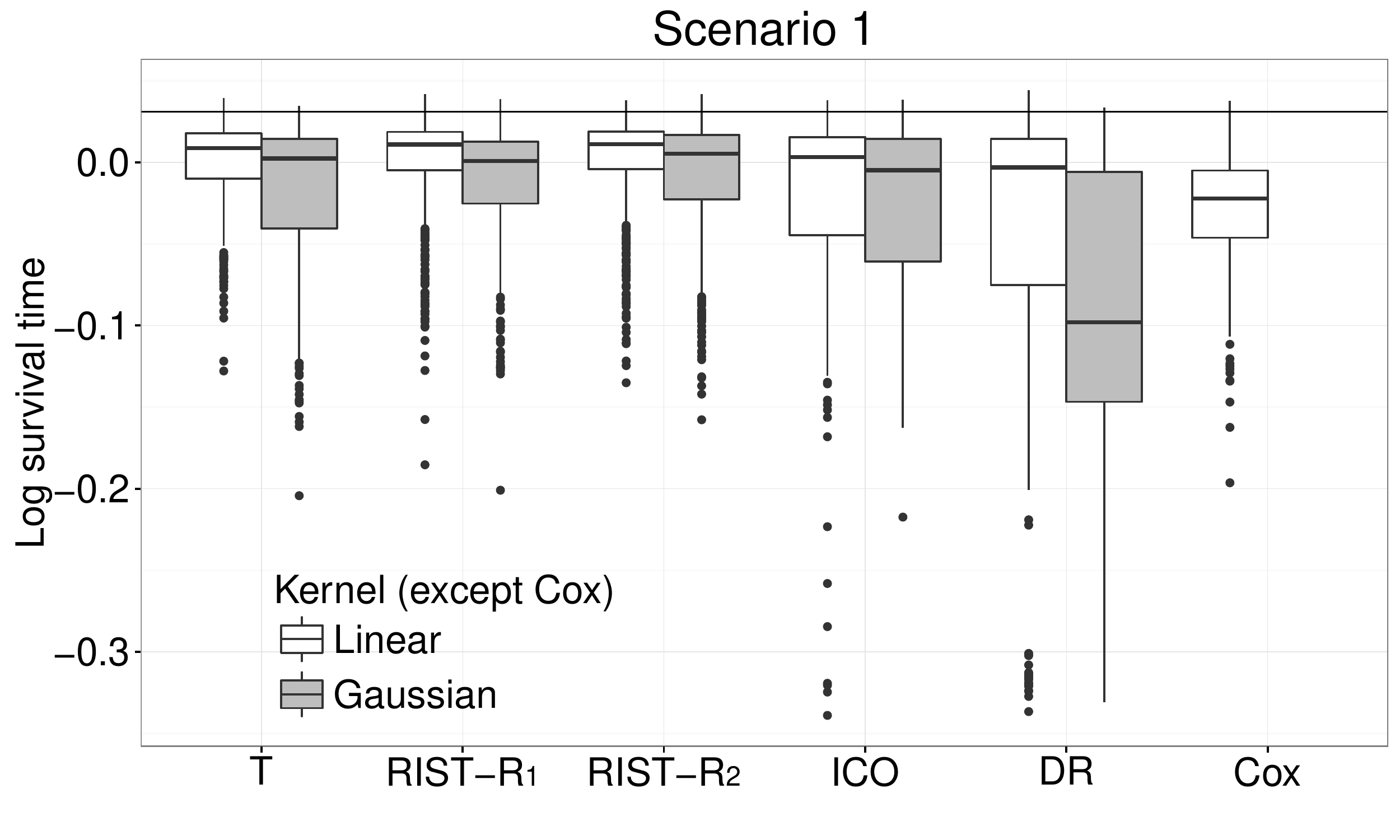} \\
   \includegraphics[trim=0in 0in 0in 0in, height=1.6in]{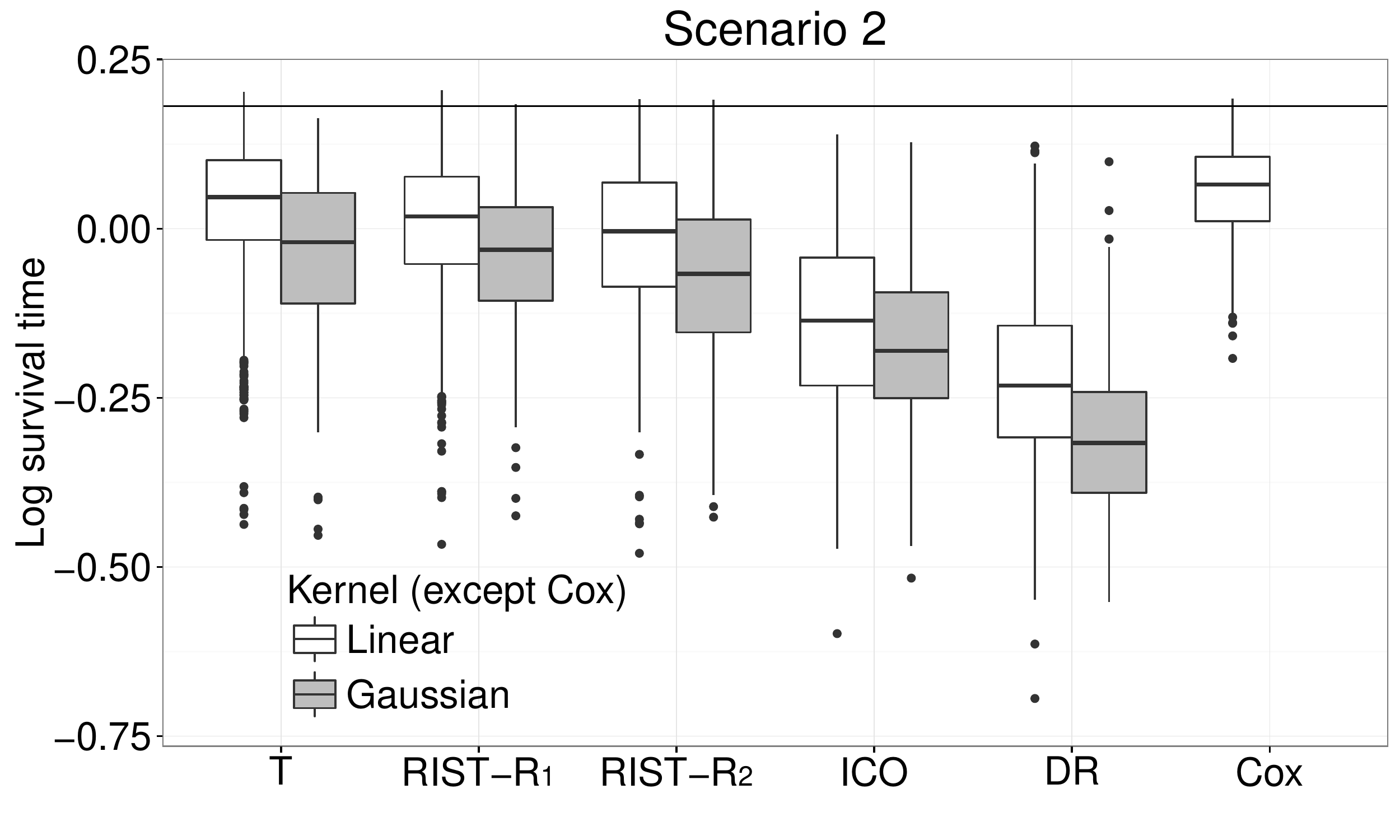} \\
   \includegraphics[trim=0in 0in 0in 0in, height=1.6in]{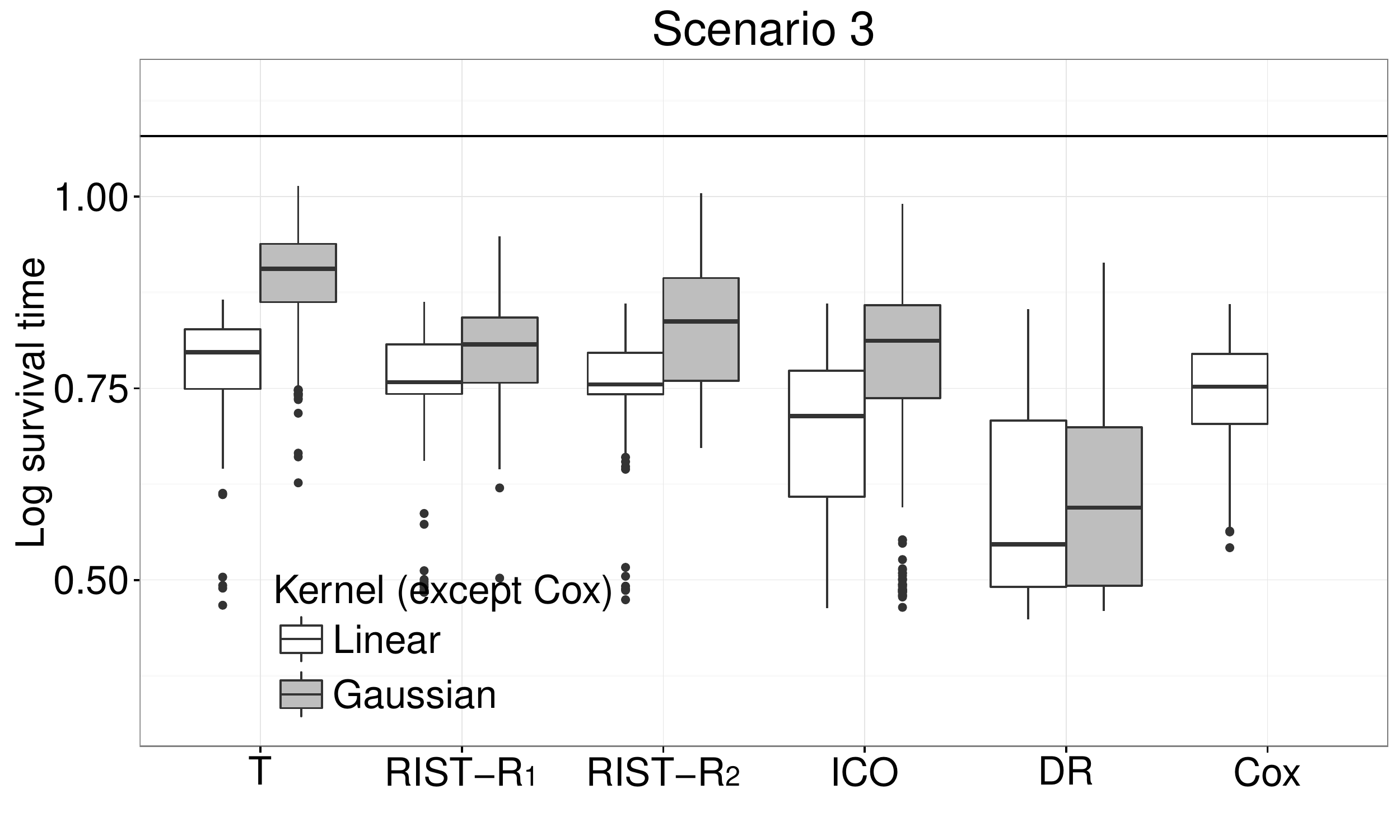} \\
   \includegraphics[trim=0in 0in 0in 0in, height=1.6in]{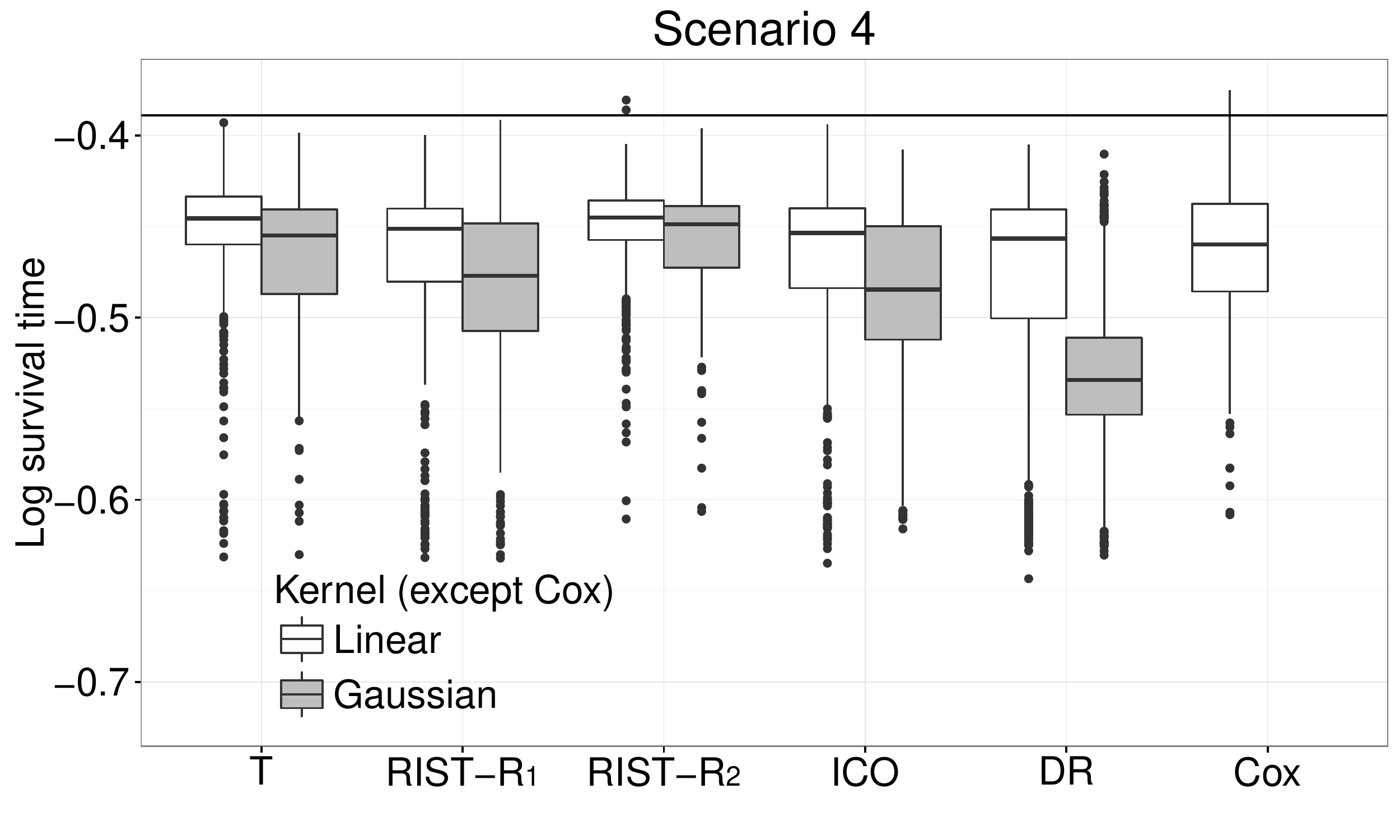} \\
    \caption{Boxplots of mean log survival time for different treatment regimes. Censoring rate: 45\%. T: using true survival time as weight; RIST-$R_1$ and RIST-$R_2$: using the estimated $R_1$ and $R_2$ respectively as weights, while the conditional expectations are estimated using recursively imputed survival trees;  ICO: inverse probability of censoring weighted learning; DR: doubly robust outcome weighted learning. The black horizontal line is the theoretical optimal value.}
    \label{fig1}
\end{figure}

Figure \ref{fig1} shows the boxplot of values based on the logarithm of $T$ calculated from the test data. The mean and standard deviation of values are shown in Table \ref{tab1}. In scenario 1, since the model is not correctly specified for inverse probability of censoring outcome weighted learning, the doubly robust estimator, or Cox regression, our method performs better than all other competitors.

\begin{table}
\caption{\label{tab1}
Simulation results: Mean ($\times 10^3$) and (sd) ($\times 10^3$). Censoring rate: 45\%. For each scenario, the theoretical optimal value ($\times 10^3$) is 31, 181, 1079, and -389, respectively.}
\begin{tabular}{cccccccc}
\noalign{\smallskip}
\noalign{\smallskip}
 & kernel & T & RIST-$R_1$ & RIST-$R_2$ & ICO & DR & Cox\\
\noalign{\smallskip}
\multirow{2}{*}{1} & Linear & ~~~0 ~(26) & ~~~0 ~(31) & ~~~1 ~(30) & ~-20 ~(54) & ~-39 ~(76) & \multirow{2}{*}{~-29 (33)}\\
                & Gaussian  & ~-17 ~(44) & ~-11 ~(35) & ~~-8 ~(36) & ~-25 ~(50) & ~-88 ~(79)\\
\noalign{\smallskip}
\multirow{2}{*}{2} & Linear & ~~22 (113) & ~~-1 (112) & ~-24 (125) & -137 (131) & -232 (132) & \multirow{2}{*}{~~53 (69)}\\
                & Gaussian  & ~-39 (115) & ~-40 (103) & ~-72 (114) & -175 (120) & -311 (106)\\
\noalign{\smallskip}
\multirow{2}{*}{3} & Linear & ~785 ~(52) & ~766 ~(59) & ~763 ~(51) & ~683 (113) & ~598 (120) & \multirow{2}{*}{~745 (64)}\\
                & Gaussian  & ~896 ~(61) & ~803 ~(56) & ~834 ~(71) & ~785 (105) & ~606 (115)\\
\noalign{\smallskip}
\multirow{2}{*}{4} & Linear & -453 ~(37) & -469 ~(47) & -451 ~(27) & -469 ~(48) & -481 ~(59) & \multirow{2}{*}{-464 (36)}\\
                & Gaussian  & -465 ~(35) & -482 ~(44) & -457 ~(28) & -487 ~(45) & -531 ~(43)\\
\noalign{\smallskip}
\end{tabular}
\source{T: using true survival time as weight; RIST-$R_1$ and RIST-$R_2$: using the estimated $R_1$ and $R_2$ respectively as weights, while the conditional expectations are estimated using recursively imputed survival trees; ICO: inverse probability of censoring weighted learning; DR: doubly robust outcome weighted learning; Cox: Cox proportional hazards model using covariate-treatment interactions.}
\end{table}

In scenario 2, we added some nonlinear terms into both the Cox and accelerated failure time models. The model assumptions for inverse censoring outcome weighted learning and the doubly robust estimator are not satisfied. Our estimated treatment rule performs much better than these two. Compared with inverse censoring outcome weighted learning and doubly robust learning, both our approaches improve more than 0.1 for the mean. Since the true model for the failure time is the Cox model, Cox regression performs better here. In this case, the Gaussian kernel performs less well than the linear kernel for most methods since the true model structure is linear and the Gaussian kernel is too flexible.

For scenario 3, which has a more complicated tree structure, the Gaussian kernel performs better than the linear kernel for all outcome weighted learning approaches. The performance of the Gaussian kernel is enhanced since it can better address the true nonlinear model structure. We can see that with either a linear or Gaussian kernel, our estimators perform better than Cox regression. Compared with doubly robust learning, our two approaches improve 0.2 for the mean.

In scenario 4, we see that when the model is correctly specified for inverse probability of censoring outcome weighted learning and doubly robust learning, the performances of both approaches are satisfactory while our methods seem to be only a little better. The performances of our first approach, inverse probability of censoring outcome weighted learning and Cox regression are all similar. Our second approach has the best treatment effect among all estimators. Note that our second approach appears to perform as well as the first, oracle approach. Also, our two proposed methods have smaller standard errors in scenarios 1 and 3. The standard error is similar for all outcome weighted learning approaches in scenario 2 and 4. Overall, our proposed methods have generally lower variances.

Compared with results of censoring rates (30\% and 60\%) in the Appendix, we can observed a consistently pattern that lower censoring rate leads to higher performances in terms of both mean value and variance. The relative performances between the proposed and the competing methods remain similar across different censoring rates.

\section{Data Analysis}\label{section5}
We apply the proposed method to a non-small-cell lung cancer randomized trial dataset described in \cite{socinski2002phase}. 228 subjects with complete information are used in this analysis. Each treatment arm contains 114 subjects. Here we use five covariates: performance status (119 subjects ranging from 90\% to 100\% and 109 subjects ranging from 70\% to 80\%), cancer stage (31 subjects in stage 3 and 197 subjects in stage 4), race (167 white, 54 black and 7 others), gender (143 male and 85 female), age (ranging from 31 to 82 with median 63). The length of study is $\tau = 104$ weeks. We adopt the same tuning parameters used in the simulation study for this analysis. The value function is again calculated by using the logarithm of survival time $\log(T)$ (in weeks) as the reward.

We randomly divide the 228 patients into four equal proportions and use three parts as training data to estimate the optimal rule and calculate the empirical value based on the remaining part. We then permute the training and testing portions and average the four results. This procedure is then repeated 100 times and averaged to obtain the mean and standard deviation. To calculate the testing data performance, we consider two different measurements, both are calculated based on the formula $\sum_{i=1}^n  R_i I\{A_i=\mathcal{D}(X_i)\}/ \sum_{i=1}^n I\{A_i=\mathcal{D}(X_i)\}$ for the testing samples, where two versions of $R_i$'s are used. We first consider the procedure proposed in \cite{zhao2015doubly}, where $R$ is defined as
{\small
\begin{align*}
\frac{\Delta Y}{\widehat S_C(Y \mid A,\bX)} -\int
\widehat E_{\widetilde T}\{T\mid T>t, A, X\}\left\{\frac{dN_{C}(t)}{\widehat
    S_C(t \mid A,\bX)}+I(Y_i\ge t) \frac{d\widehat S_C(t \mid A,\bX)}{\widehat
    S_C(t \mid A,\bX)^2}\right\}.
\end{align*}
}
Here, $\widehat S_C(t \mid A,\bX)$ and $\widehat E_{\widetilde T}(T\mid T>t,A,X)$ are estimated from the Cox model for simplicity. We also consider a more direct clinical measurement without the double robustness correction, which can be interpreted in a similar way as the expected survival time or the restricted mean survival time \cite{geng2015optimal,ma2015statistical,tian2014predicting}. To be specific, we consider a restricted mean (log) survival time truncated at $\tau$ defined as $\delta T +(1-\delta)E(T)$, and use this as a plug-in quantity of $R$ in the testing performance calculation. To estimate this quantity, we use a recursively imputed survival trees (RIST) method to produce the expected survival time $E(T)$. 

\begin{figure}[H]
\centering
   \includegraphics[trim=0in 0in 0in 0in, height=2in]{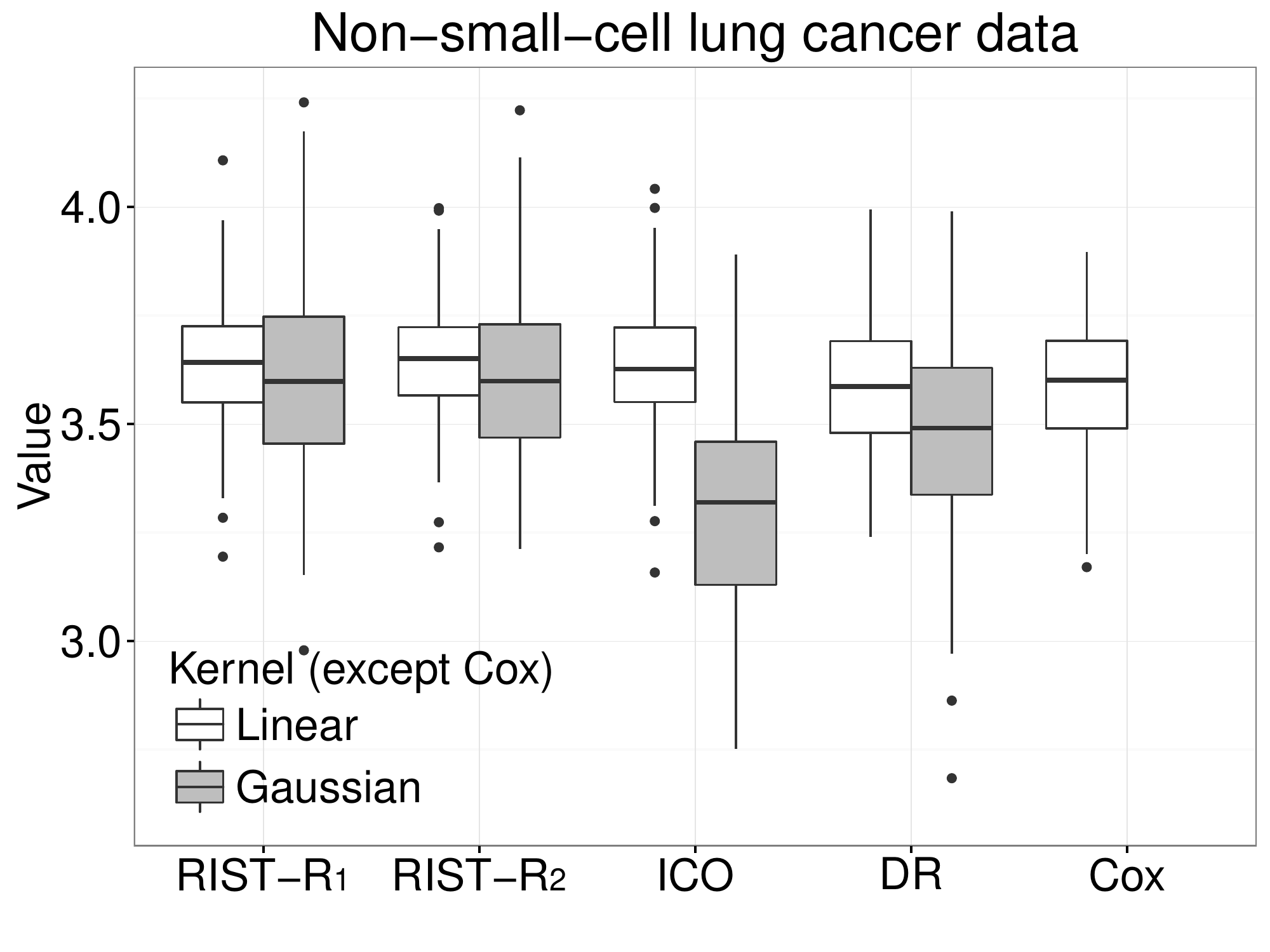}
    \caption{Boxplots of cross-validated value of survival weeks on the log scale. RIST-$R_1$ and RIST-$R_2$: using the estimated $R_1$ and $R_2$ respectively as weights, while the conditional expectations are estimated using recursively imputed survival trees; ICO: inverse probability of censoring weighted learning; DR: doubly robust outcome weighted learning.}
    \label{fig2}
\end{figure}

The value function results are presented in Table \ref{tab2} and Figure \ref{fig2}. Both proposed methods have higher values than the compared methods. Note that for the Gaussian kernel, our two new approaches are still better than Cox regression, however, inverse probability of censoring outcome weighted learning and doubly robust learning are not much different from Cox regression. The standard error is comparable among all four methods using the linear kernel. For the Gaussian kernel, the standard errors of the proposed methods and inverse probability of censoring weighted learning are similar. The standard error for the doubly robust method is slightly worse in this instance. Overall, the proposed methods seem to perform best.

\begin{table}[H]
\caption{\label{tab2} Analysis of non-small-cell lung cancer data: Mean (sd) of value function}
\begin{tabular}{ccccccc}
\noalign{\smallskip}
\noalign{\smallskip}
kernel & RIST-$R_1$ & RIST-$R_2$ & ICO & DR & Cox\\
\noalign{\smallskip}
Linear      & 3.641 (0.144) & 3.641 (0.138) & 3.633 (0.158) & 3.590 (0.174) & \multirow{2}{*}{3.582 (0.158)}\\
Gaussian    & 3.611 (0.215) & 3.615 (0.220) & 3.302 (0.221) & 3.470 (0.233)\\
\noalign{\smallskip}
\noalign{\smallskip}
\end{tabular}
\source{RIST-$R_1$ and RIST-$R_2$: using the estimated $R_1$ and $R_2$ respectively as weights, while the conditional expectations are estimated using recursively imputed survival trees; ICO: inverse probability of censoring weighted learning; DR: doubly robust outcome weighted learning; Cox: Cox proportional hazards model using covariate-treatment interactions.}
\end{table}

\begin{figure}[H]
\centering
   \includegraphics[trim=0in 0in 0in 0in, height=2in]{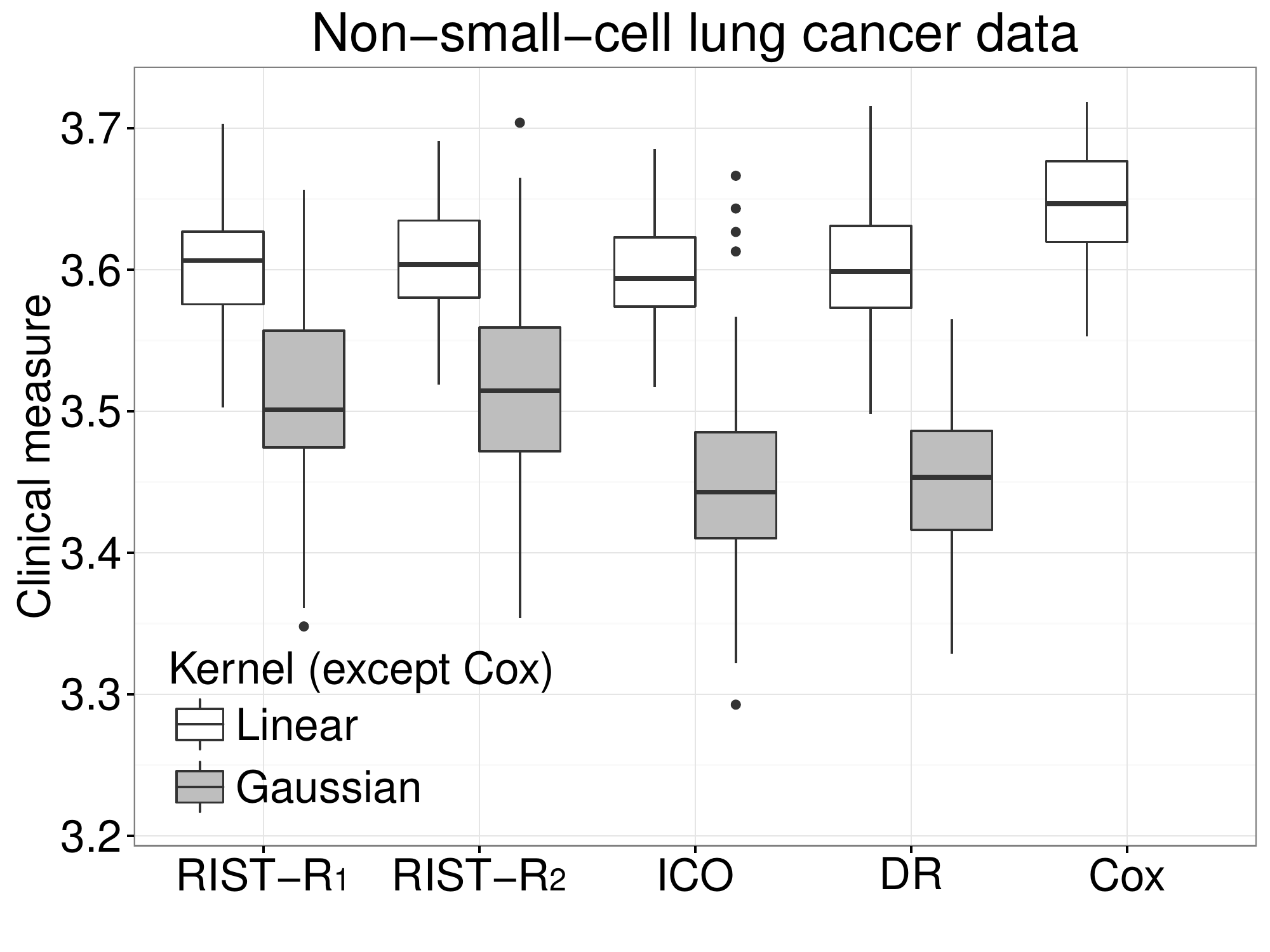}
    \caption{Boxplots of cross-validated value of survival weeks on the log scale. RIST-$R_1$ and RIST-$R_2$: using the estimated $R_1$ and $R_2$ respectively as weights, while the conditional expectations are estimated using recursively imputed survival trees; ICO: inverse probability of censoring weighted learning; DR: doubly robust outcome weighted learning.}
    \label{fig5}
\end{figure}

The restricted log mean results are presented in Table \ref{tab5} and Figure \ref{fig5}. Note for the linear kernel, the median of the proposed methods are higher than 3.6 and median of both inverse probability of censoring outcome weighted learning and doubly robust learning are lower. For the Gaussian kernel, the proposed methods are much better than inverse probability of censoring outcome weighted learning and doubly robust learning. Interestingly, under this measure, the performance of Cox regression is the best. A possible reason is that the true underlying model may not deviate much from the proportional hazard model, making the Cox model a better choice. This is also reflected by the fact that the results look similar to the simulation Scenario 2 plot, where the Cox model performs the best. Another possible reason is that the pseudo-outcome estimated from RIST may not be completely accurate and favors the Cox model in this particular dataset.

\begin{table}[H]
\caption{\label{tab5} Analysis of non-small-cell lung cancer data: Mean (sd) of a clinical measure}
\begin{tabular}{ccccccc}
\noalign{\smallskip}
\noalign{\smallskip}
kernel & RIST-$R_1$ & RIST-$R_2$ & ICO & DR & Cox\\
\noalign{\smallskip}
Linear      & 3.603 (0.040) & 3.606 (0.037) & 3.598 (0.037) & 3.601 (0.042) & \multirow{2}{*}{3.646 (0.039)}\\
Gaussian    & 3.511 (0.064) & 3.514 (0.068) & 3.451 (0.062) & 3.456 (0.052)\\
\noalign{\smallskip}
\noalign{\smallskip}
\end{tabular}
\source{RIST-$R_1$ and RIST-$R_2$: using the estimated $R_1$ and $R_2$ respectively as weights, while the conditional expectations are estimated using recursively imputed survival trees; ICO: inverse probability of censoring weighted learning; DR: doubly robust outcome weighted learning; Cox: Cox proportional hazards model using covariate-treatment interactions.}
\end{table}

\section{Discussion}\label{section6}
We proposed a new method that redefines the reward function in a censored survival setting. The method works by replacing the censored observations (or all observations) by an estimated conditional expectation of the failure time. In practice, the failure time (or logarithm of the failure time) is commonly used in defining the reward function $R$, however, this choice could more flexible. For example, we may be interested in searching for a treatment rule that maximizes the median survival time or a certain quantile. Under our framework, this is achievable by replacing the censored observations with a suitable estimate of the quantile. This part of the work is currently under investigation.

The proposed methods may be improved or extended in multiple ways. The estimated treatment rule may be affected by the shift of the outcome. A potential extension is to combine our methods with residual weighted learning \citep{zhou2015residual}, which has been shown to reduce the total variation of the weights and improve stability. Trials with multiple treatment arms occur frequently. Thus a potential extension of our method is in the direction of multicategory classification \citep{bredensteiner1999multicategory,lee2004multicategory}. It is also interesting to extend our method to dynamic treatment regimes where a sequence of decision rules \citep{murphy2003optimal,zhao2011reinforcement,laber2014dynamic,zhao2015new} need to be learned in a censored survival outcome setting \citep{goldberg2012q}.

\section*{Acknowledgment}
This research is supported in part by U.S. National Science Foundation grant DMS-1407732 and by U.S. National Institutes of Health grant P01 CA142538. We thank Yinqi Zhao for helpful conversations and suggestions. We thank the editor, associated editor, and reviewers for helpful comments which led to an improved manuscript.

\appendix

\section*{Appendix}
\subsection*{A simplified tree-based survival model used in Theorem \ref{tree} \label{app-treecons}}

We consider a simplified version of a tree-based survival model. Starting from the root node $[0, 1]^d$, at each internal node, we randomly chose the $j$-th feature of $X$ to split the node, while the splitting point is always at the midpoint of the range of the chosen feature. We repeat splitting $\lceil \log_2 k_n \rceil$ times, where $k_n$ is a deterministic parameter which we can control. Hence, each individual tree has exactly $2^{\left \lceil{\log_2 k_n}\right \rceil }$ terminal nodes, which is approximately $k_n$. In practice, we always chose $k_n$ to go to infinity as $n$ goes to infinity.

After we build an individual tree, let $B_i~(i=1,2,\ldots, 2^{\left \lceil{\log_2 k_n}\right \rceil })$ be the rectangular cell of the random partition. We treat observations inside each leaf node as a group of homogeneous subjects and compute the Nelson-Aalen estimator $\hat \Lambda(\cdot \mid B_i)$ for each leaf node $B_i$. Hence, our estimator is essentially
\begin{align*}
\hat r_n(\cdot,X,A) =\sum_{i=1}^{2^{\left \lceil{\log_2 k_n}\right \rceil }} I\{(X,A)\in B_i\}\hat \Lambda(\cdot \mid B_i).
\end{align*}

\subsection*{Proof of Theorem \ref{tree}}
\begin{proof}
Since we always assume that the treatment variable $A$ is important, and $A$ has only two categories, we force a split on $A$ at the root node. This is equivalent to fitting trees for $A=1$ and $A=-1$ separately. In a balanced design, the problem reduces to estimating $r(\cdot, X, 1)$ or $r(\cdot, X, -1)$ with sample size $n/2$. Without the risk of ambiguities, the following results are developed for $\hat r_n(\cdot, X)$ with sample size $n$, where the results can be applied to either $A=1$ or $-1$. Our proof utilizes two facts from \citep{biau2012analysis}:

\textbf{Fact 1}
Let $K_{nj}\{B_i\}$ be the number of times the $j$-th coordinate ($j=1,\ldots,d$) is split on to reach the terminal node $B_i, (i=1,2,\ldots, 2^{\left \lceil{\log_2 k_n}\right \rceil })$. Conditionally on $X$,  $K_{nj}\{B_i\}$ is $Binomial(\left \lceil{\log_2 k_n}\right \rceil, 1/d)$. Moreover, $\sum_{j=1}^d K_{nj}\{B_i\}=\left \lceil{\log_2 k_n}\right \rceil$.

\textbf{Fact 2}
Let $N_n(B_i)$ be the number of data points falling in the cell $B_i, (i=1,2,\ldots, 2^{\left \lceil{\log_2 k_n}\right \rceil })$. Conditionally on $\Theta$, $N_n(B_i)$ follows $Binomial(n,2^{-\left \lceil{\log_2 k_n}\right \rceil})$.

The following lemma, for later reference, provides the deterministic limit of the Nelson-Aalen estimator in the independent non-identically distributed case. The proof can be found in an unpublished technical report by Mai Zhou at the University of Kentucky.

\begin{lemma}

Suppose we have two sets of non-negative random variables: \\ $T_1,T_2,\ldots,T_n$ which are survival times, independent but non-identically distributed with continuous distribution $F_1(t),F_2(t),\ldots,F_n(t)$; $C_1,C_2,\ldots,C_n$\\ which are censoring times, independent but non-identically distributed with continuous distribution $G_1(t),G_2(t),\ldots,G_n(t)$. We also assume the $T_i's$ and $C_i's$ are independent. The Nelson-Aalen estimator of data $Y_i=\min(T_i,C_i),\delta_i=I(T_i\leq C_i)$ is $\hat \Lambda(t)$. Provided Assumption \ref{addas}, we have
\begin{align}\label{NA}
\text{pr}(\sup_{t<\tau} |\hat \Lambda(t)- \int_0^t \frac{\sum_i \{1-G_i(s)\}dF_i(s)}{\sum_i\{1-G_i(s)\}\{1-F_i(s)\}} | > \frac{(1152b)^{1/2}}{n^{1/2}M^2})< 16(n+2)e^{-b},
\end{align}
where $b>1/228$, $n \geq 288b/M^4$.
\end{lemma}

Now we start the proof of Theorem \ref{tree}. Let the limit of the Nelson-Aalen estimator inside the cell $B_i~(i=1,2,\ldots, 2^{\left \lceil{\log_2 k_n}\right \rceil })$ be
\begin{align*}
\Lambda^*(t\mid B_i)=\int_0^t \frac{[\sum_{\bX_j\in B_i}\{1-G_j(s)\}dF_j(s)]}{[\sum_{\bX_j\in B_i} \{1-G_j(s)\}\{1-F_j(s)\}]}.
\end{align*}
For any $t<\tau$, in order to bound the $|\hat r_n(t, \bX)-r(t, \bX)|$, we define
\begin{align*}
r^*_n(t, \bX) & =\sum_{i=1}^{ 2^{\left \lceil{\log_2 k_n}\right \rceil }} I\{X\in B_i\} \Lambda^*(t \mid B_i).
\end{align*}
Then $|\hat r_n(t, \bX)-r(t, \bX)|$ can be decomposed as
\begin{align}\label{six}
|\hat r_n(t, \bX)-r(t, \bX)| =|\hat r_n(t, \bX)- r^*_n(t, \bX)|+|r^*_n(t, \bX)-r(t, \bX)|.
\end{align}
We start with the first term in Equation \eqref{six}. From Fact 2, we know the number of observations in each terminal node is $Binomial(n,2^{-\lceil \log_2 k_n \rceil})$. By the Chernoff bound, with probability larger than $1-e^{-u^2 n 2^{-\lceil \log_2 k_n \rceil-1}}$, in one terminal node we have at least $(1-u)n 2^{-\lceil \log_2 k_n \rceil}$ observations for some $0<u<1$.

Combining Equation \eqref{NA}, the following equation holds:
\begin{align}\label{com1}
|\hat r_n(t, \bX)-& r^*_n(t, \bX)| \nonumber \\
\leq& \sum_{i=1}^{ 2^{\left \lceil{\log_2 k_n}\right \rceil }}  I\{X\in B_i\}(1152b)^{1/2}\{(1-u)n 2^{-\lceil \log_2 k_n \rceil}\}^{-1/2}M^{-2} \nonumber \\
=&~(1152b)^{1/2}\{(1-u)n 2^{-\lceil \log_2 k_n \rceil}\}^{-1/2}M^{-2},
\end{align}

with probability $1-16[(1-u)n 2^{-\lceil \log_2 k_n \rceil}+2]e^{-b}-e^{-u^2 n 2^{-\lceil \log_2 k_n \rceil-1}}$, where $b>1/228$ and $(1-u)n 2^{-\lceil \log_2 k_n \rceil}\geq 288b/M^4$.

Before we bound the second term in Equation \eqref{six}. We first show the bound for the difference between the true cumulative hazard function and aggregated estimator inside the cell $B_i~ (i=1,2,\ldots, 2^{\left \lceil{\log_2 k_n}\right \rceil }) $, i.e. $|I\{X\in B_i\}\{ \Lambda^*(t \mid B_i)-\Lambda(t \mid \bX)\}|$.

From Fact 1, we know the number of times the terminal node $B_i$ is split on the $j$-th coordinate ($j=1,\cdots,d$) $K_{nj}\{B_i\}$ is $Binomial(\left \lceil{\log_2 k_n}\right \rceil ,1/d)$. By the Chernoff bound, $P(K_{nj}\{B_i\}\leq (1-r)\left \lceil{\log_2 k_n}\right \rceil /d) \leq e^{-\left \lceil{\log_2 k_n}\right \rceil  r^2/(2d)}$ for some $0<r<1$. So with probability $(1-e^{-\left \lceil{\log_2 k_n}\right \rceil r^2/(2d)})^d \geq 1-de^{-\left \lceil{\log_2 k_n}\right \rceil  r^2/(2d)}$, every dimension of $B_i$ is less than $2^{-\{(1-r)\left \lceil{\log_2 k_n}\right \rceil \}/d}$. Then we have with probability larger than $1-de^{-\left \lceil{\log_2 k_n}\right \rceil  r^2/(2d)}$ we have
 \begin{align*}
\max_{\bX_1,\bX_2\in B_i}||\bX_1-\bX_2||\leq d^{1/2}2^{-\{(1-r)\left \lceil{\log_2 k_n}\right \rceil \}/d}.
 \end{align*}
 So for all the observations $X_j$ inside the same cell as $X$, by Assumption \ref{LC}, we have
 \begin{align*}
 |F_X(\cdot)- F_j(\cdot)| \leq L d^{1/2}2^{-\{(1-r)\left \lceil{\log_2 k_n}\right \rceil \}/d},
  \end{align*}
   \begin{align*}
 |f_X(\cdot)-f_j(\cdot)|  \leq (L'+L^2) d^{1/2}2^{-\{(1-r)\left \lceil{\log_2 k_n}\right \rceil \}/d},
   \end{align*}
where $f_X(\cdot)$ and $F_X(\cdot)$ denote the true density function and distribution function at $X$, respectively. Then $\Lambda^*(t\mid B_i)$ has the upper bound and lower bound
\begin{align*}
\int_0^t  [f_X(s)+b_1]/[1-F_X(s)-b_2]ds \quad \text{and} \quad \int_0^t  [f_X(s)-b_1]/[1-F_X(s)+b_2]ds,
\end{align*}
respectively, where
\begin{align*}
b_1 &=(L'+L^2) d^{1/2}2^{-\{(1-r)\left \lceil{\log_2 k_n}\right \rceil \}/d} \quad \text{and} \quad b_2 =Ld^{1/2}2^{-\{(1-r)\left \lceil{\log_2 k_n}\right \rceil \}/d}.
\end{align*}
Hence, $|I\{X\in B_i\}\{ \Lambda^*(t \mid B_i)-\Lambda(t\mid \bX)\}|$ has the bound
\begin{align*}
\int_0^t \frac{b_1 (1-F(s)) +b_2 f(s)}{(1-F(s)-b_2)(1-F(s))}ds
\leq C \tau d^{1/2}2^{-\{(1-r)\left \lceil{\log_2 k_n}\right \rceil \}/d},
\end{align*}
where $C$ is some constant depending on $L$ and $L'$. We then bound the second term of Equation \eqref{six} as follows:
\begin{align}\label{com2}
\begin{split}
|r^*_n(t, \bX)-r(t, \bX)| & \leq \sum_{i=1}^{ 2^{\left \lceil{\log_2 k_n}\right \rceil }} I\{X\in B_i\} | \Lambda^*(t \mid B_i)-\Lambda(t\mid \bX)|\\
& \leq C \tau d^{1/2}2^{-\{(1-r)\left \lceil{\log_2 k_n}\right \rceil \}/d}.
\end{split}
\end{align}

Combining Equation \eqref{com1} and \eqref{com2}, For each $X$, we have
\begin{align*}
\text{pr}[\sup_{t<\tau} |\hat r_n(t, \bX)&-r(t, \bX)|\leq C[ \tau d^{1/2}2^{-\{(1-r)\left \lceil{\log_2 k_n}\right \rceil \}/d}\\
&+(1152b)^{1/2}\{(1-u)n 2^{-\lceil \log_2 k_n \rceil}\}^{-1/2}M^{-2} ] \geq  1-w_n,
\end{align*}
where
\begin{align*}
w_n=~ 16[(1-u)n 2^{-\lceil \log_2 k_n \rceil}+2]e^{-b} + e^{-u^2 n 2^{-\lceil \log_2 k_n \rceil-1}} + de^{-\left \lceil{\log_2 k_n}\right \rceil  r^2/(2d)}.
\end{align*}
This completes the proof.
\end{proof}

\subsection*{Proof of Theorem \ref{tree2}}

\begin{proof}
Based on Theorem \ref{tree}, we now only need to establish the bound of \\$|\hat r_n(t,X,A)-r(t,X,A)|$ under the event with small probability $w_n$. Noticing that $\hat r_n(t,X,A)$ is simply the Nelson-Aalen estimator of the cumulative hazard function with at most $n$ terms, for any $t<\tau$ we have
\begin{align*}
 \hat r_n(t,X,A)\leq \frac{1}{n}+\ldots+\frac{1}{1}=O(\ln(n)),
 \end{align*}
which implies that
\begin{align*}
|\hat r_n(t,X,A)-r(t,X,A)|\leq O(\ln(n)).
\end{align*}
Combining this with Theorem \ref{tree} completes the proof.
\end{proof}

\subsection*{Proof of Lemma \ref{lemma1}}
\begin{proof}
Our survival function estimator is $\hat S(t)=e^{-\hat \Lambda(t)}$. From Theorem \ref{tree}, we know that for any $t< \tau$,
{\small
\begin{align*}
 &pr(|\hat S(t\mid X,A) -S(t\mid X,A) |\leq C[2^{-(1-r)\left \lceil{\log_2 k_n}\right \rceil/d}+(b/\{(1-u)n 2^{-\lceil \log_2 k_n \rceil}\})^{1/2}] )\\
& \geq 1-16[(1-u)n 2^{-\lceil \log_2 k_n \rceil}+2]e^{-b}-e^{-u^2 n 2^{-\lceil \log_2 k_n \rceil-1}}-de^{-\left \lceil{\log_2 k_n}\right \rceil r^2/(2d)}.
\end{align*}
}
It is then easy to see that for $R_1$,
\begin{align*}
&\Big| \widehat E(T \mid X,A) -E(T \mid X,A) \Big| \\
=& ~\Big| \int_0^\tau \widehat S(t \mid X,A)dt - \int_0^\tau S(t \mid X,A)dt \Big| \\
\leq& \int_0^\tau| \widehat S(t \mid X,A) - S(t \mid X,A)|dt \\
\leq& ~\tau C[2^{-\{(1-r)\left \lceil{\log_2 k_n}\right \rceil \}/d}+(b/\{(1-u)n 2^{-\lceil \log_2 k_n \rceil}\})^{1/2}],
\end{align*}
with probability larger than $1-w_n$. And for reward $R_2$, we have
\begin{align*}
&~ \big| \widehat E(T \mid X,A,T>Y,Y) -E(T \mid X,A,T>Y,Y) \big| \\
=& ~\Big| \int_Y^\tau \{\widehat S(t \mid X,A)/\widehat S(Y \mid X,A)\} dt - \int_Y^\tau \{ S(t \mid X,A) / S(Y \mid X,A)\}dt \Big| \\
\leq& ~\Big| \int_Y^\tau \{\widehat S(t \mid X,A)/\widehat S(Y \mid X,A)\} dt - \int_Y^\tau \{\widehat S(t \mid X,A) / S(Y \mid X,A)\}dt \Big| \\
+& ~\Big| \int_Y^\tau \{\widehat S(t \mid X,A)/ S(Y \mid X,A)\} dt - \int_Y^\tau \{ S(t \mid X,A) / S(Y \mid X,A)\}dt \Big|.
\end{align*}
Note that we can bound the distance between $\widehat S(Y \mid X,A)$ and $S(Y \mid X,A)$ with probability no less than $1-w_n$, which is further bounded above by
\begin{align*}
(1/M^2 + 1/M) & \int_Y^\tau| \widehat S(Y \mid X,A) - S(Y \mid X,A)|dt \\
&\leq C_2 [2^{-\{(1-r)\left \lceil{\log_2 k_n}\right \rceil \}/d}+(b/\{(1-u)n 2^{-\lceil \log_2 k_n \rceil}\})^{1/2}],
\end{align*}
for some constant $C_2$ with probability larger than $1-2w_n$.
\end{proof}

\subsection*{Proof of Theorem \ref{thm3}}
\begin{proof}
We restate the value function corresponding to the true and working model as
\begin{align*}
V(f) &= E( RI[A = \sign\{f(X)\}]/\pi(A;X)) \\
\text{and} \,\,\, V'(f) &= E(\widehat RI[A = \sign\{f(X)\}]/\pi(A;X)),
\end{align*}
respectively. Then we have
\begin{align}
V(f^*)-V(\widehat f_n)\leq& ~V(f^*)-\sup_{f \in \mathcal{F}} V'(f)+\sup_{f \in \mathcal{F}} V'(f)-V'(\widehat f_n)+V'(\widehat f_n)-V(\widehat f_n) \nonumber\\
\leq& ~V(f^*)-V'(f^*)+\sup_{f \in \mathcal{F}} V'(f)-V'(\widehat f_n)+V'(\widehat f_n)-V(\widehat f_n) \nonumber\\
\leq& ~\sup_{f \in \mathcal{F}} V'(f)-V'(\widehat f_n)+2\sup_{f \in \mathcal{F}}|V(f)-V'(f)|. \label{eq00}
\end{align}
We start with the first term in Equation (\ref{eq00}). From Lemma \ref{lemma0}, we know that $\sup_{f \in \mathcal{F}} V'(f)-V'(\widehat f_n)=V'(\widetilde f)-V'(\widehat f_n)$, where $\widetilde f=\arg\min_{f\in \mathcal{H}_k} E\{L_{\phi}(f)\}$.

Let $\widetilde f_{\lambda_n}= \arg \min_{f\in \mathcal{H}_k} [E\{{R\phi\{Af(X)\}}/{\pi(A;X)}\}+\lambda_n\|f\|^2_k]$, then
\begin{align}\label{eq0}
n^{-1}\sum_{i=1}^n\frac{\widehat R\phi\{A_i\widehat f(\bX_i)\}}{\pi(A_i;X_i)} +\lambda_n\|\widehat{f}\|_k^2 \leq n^{-1}\sum_{i=1}^n\frac{\widehat R\phi\{A_i\widetilde f_{\lambda_n}(X_i)\}}{\pi(A_i;X_i)}+\lambda_n\|\widetilde f_{\lambda_n}\|_k^2.
\end{align}
By the definition of $a(\lambda)$, we have
\begin{align*}
a(\lambda_n)=[ E\{L_{\phi}(\widetilde f_{\lambda_n})\}+\lambda_n ||\widetilde f_{\lambda_n}||_k^2 -E\{L_{\phi}(\widetilde f)\}],
\end{align*}
and by Theorem 3.2 in \citep{zhao2012estimating}, we further have
\begin{align*}
V'(\tilde f) - V'(\widehat{f}) \leq& ~E\{L_{\phi}(\widehat f)\}-E\{L_{\phi}(\widetilde f)\}\\
\leq& ~E\{L_{\phi}(\widehat f)\}-E\{L_{\phi}(\widetilde f_{\lambda_n})\}-\lambda_n||\widetilde f_{\lambda_n}||_k^2 \\
+& E\{L_{\phi}(\widetilde f_{\lambda_n})\}-E\{L_{\phi}(\widetilde f)\}+\lambda_n ||\widetilde f_{\lambda_n}||_k^2\\
\leq& ~E\{L_{\phi}(\widehat f)\}-E\{L_{\phi}(\widetilde f_{\lambda_n})\}-\lambda_n||\widetilde f_{\lambda_n}||_k^2+\lambda_n||\widehat f||_k^2+a(\lambda_n).
\end{align*}
Combined with \eqref{eq0},
\begin{align*}
V'(\widetilde f)-& V'(\widehat{f})
\leq ~a(\lambda_n) + E\left[\frac {R\phi\{A\widehat{f} (X)\}}{\pi(A;X)} - \frac {\widehat R\phi\{A\widehat{f}(X)\}}{\pi(A;X)}\right] \\
+ &E\left[\frac {\widehat R\phi\{A \widetilde f_{\lambda_n}(X)\}}{\pi(A;X)}-\frac {R\phi\{A \widetilde f_{\lambda_n}(X)\}}{\pi(A;X)} \right]\\
+& \bigg( -n^{-1}\sum_{i=1}^n \Big[ \lambda_n\|\widehat f\|_k^2 + \frac {\widehat R\phi\{A_i\widehat{f}(X_i)\}}{\pi(A_i;\bX_i)} -\lambda_n\|\widetilde f_{\lambda_n}\|_k^2 -\frac {\widehat R\phi\{A_i \widetilde f_{\lambda_n}(X_i)\}}{\pi(A_i;X_i)} \Big]\\
+& E\Big[ \lambda_n\|\widehat f\|_k^2+ \frac {\widehat R\phi\{A\widehat{f}(X)\}}{\pi(A;X)} -\lambda_n\|\widetilde f_{\lambda_n}\|_k^2-\frac {\widehat R\phi\{A\widetilde f_{\lambda_n}(X)\}}{\pi(A;X)} \Big] \bigg)\\
=& ~ a(\lambda_n)+(\text{I})+(\text{II})+(\text{III}).
\end{align*}
Since $$n^{-1}\sum_{i=1}^n\frac{\widehat R\phi\{A_i\widehat f(\bX_i)\}}{\pi(A_i;\bX_i)} + \lambda_n\|\widehat{f}\|_k^2 \leq n^{-1}\sum_{i=1}^n\frac{\widehat R\phi(0)}{\pi(A_i;\bX_i)}=n^{-1}\sum_{i=1}^n\frac{\widehat R}{\pi(A_i;\bX_i)},$$
and the estimated value function $\widehat R$ is bounded by $\tau$, we know that $\|\widehat{f}\|_k \leq \tau^{1/2}\lambda_n^{-1/2}$. Furthermore, since
$$\lambda_n\|\widetilde f_{\lambda_n}\|_k^2  \leq \inf_{f \in \mathcal{H}_k}\lambda_n \|f\|_k^2 + E\left[\frac {R\phi\{Af(\bX)\}}{\pi(A;\bX)}\right] \leq E\left[\frac {R\phi(0)}{\pi(A;\bX)}\right],$$
we have $\|\widetilde f_{\lambda_n}\|_k \leq \tau^{1/2} \lambda_n^{-1/2}$. Combining with Lemma \ref{lemma1}, (I) and (II) are bounded by $C_1\lambda_n^{-1/2}\{2^{-\{(1-r)\left \lceil{\log_2 k_n}\right \rceil \}/d}+(b/\{(1-u)n 2^{-\lceil \log_2 k_n \rceil}\})^{1/2}+w_n \ln n\}$ for both $R_1$ and $R_2$, where $C_1$ is some constant.
 Following the results in \citep{zhao2015doubly}, (III) is bounded by
$M_v(n\lambda_n/c_n)^{-2/(v+2)}+M_v\lambda_n^{-1/2}(c_n/n)^{2/(d+2)}+K \rho(n\lambda_n)^{-1}+2K\rho n^{-1}\lambda_n^{-1/2}$ with probability larger than $1-2e^{-\rho}$, where $M_v$ is a constant depending on $v$ and $K$ is a sufficiently large positive constant. Finally, combining (I), (II) and (III), we have
\begin{align}\label{eq1}
\begin{split}
\text{pr}(\sup_{f \in \mathcal{F}} V'(f) \leq  V'(\widehat{f}_n)+\epsilon_1)& \geq  1-2e^{-\rho},
\end{split}
\end{align}
where $\epsilon_1=~a(\lambda_n)+M_v (n\lambda_n/c_n)^{-2/(v+2)}+M_v\lambda_n^{-1/2}(c_n/n)^{2/(d+2)}+K \rho(n\lambda_n)^{-1}+2K\rho n^{-1}\lambda_n^{-1/2}+C_1\lambda_n^{-1/2}\{2^{-\{(1-r)\left \lceil{\log_2 k_n}\right \rceil \}/d}+(b/\{(1-u)n 2^{-\lceil \log_2 k_n \rceil}\})^{1/2}+w_n\ln n\}$.

For the second part in Equation (\ref{eq00}),
\begin{align*}
V(f)-V'(f)=& ~E \Big(\frac{RI[A =\sign\{f(X)\}]}{\pi(A;X)}\Big)-E \Big(\frac{\widehat RI[A =\sign\{f(X)\}]}{\pi(A;X)}\Big)\\
=& ~E \Big(\{E(T \mid X,A)-\widehat E(T \mid X,A)\}\frac{I[A = \sign\{f(X)\}]}{\pi(A;X)}\Big)
\end{align*}
if $R = R_1$. For $R = R_2$, we have
{\small
\begin{align*}
&V(f)-V'(f)\\
=&~E \Big((1-\delta) \{E(T \mid X,A,T>Y,Y)-\widehat E(T \mid X,A,T>Y,Y)\}\frac{I[A =\sign\{f(X)\}]}{\pi(A;X)}\Big).
\end{align*}
}
By Lemma \ref{lemma1},
\begin{align}\label{eq2}
\begin{split}
&\sup_{f \in \mathcal{F}} |V(f)-V'(f)| \\
\leq &C_2\lambda_n^{-1/2}\{2^{-\{(1-r)\left \lceil{\log_2 k_n}\right \rceil \}/d}+(b/\{(1-u)n 2^{-\lceil \log_2 k_n \rceil}\})^{1/2}
+w_n \ln n\},
\end{split}
\end{align}
where $C_2$ is some constant. Now, combining \eqref{eq1} and \eqref{eq2} we have
\begin{align*}
\text{pr}(V(f^*) \leq V(\widehat{f}_n)+\epsilon) & \geq 1-2e^{-\rho},
\end{align*}
where
{\small
\begin{align*}
\epsilon&=~a(\lambda_n)+M_v (n\lambda_n/c_n)^{-2/(v+2)}+M_v\lambda_n^{-1/2}(c_n/n)^{2/(d+2)}+K \rho(n\lambda_n)^{-1}\\
&+2K\rho n^{-1}\lambda_n^{-1/2}+C\lambda_n^{-1/2}\{2^{-\{(1-r)\left \lceil{\log_2 k_n}\right \rceil \}/d}+(b/\{(1-u)n 2^{-\lceil \log_2 k_n \rceil}\})^{1/2}\\
&+w_n \ln n\}.
\end{align*}
}
This completes the proof.
\end{proof}

\subsection*{Additional simulation results for different censoring rates}

\begin{figure}[H]
\centering
   \includegraphics[trim=0in 0in 0in 0in, height=1.6in]{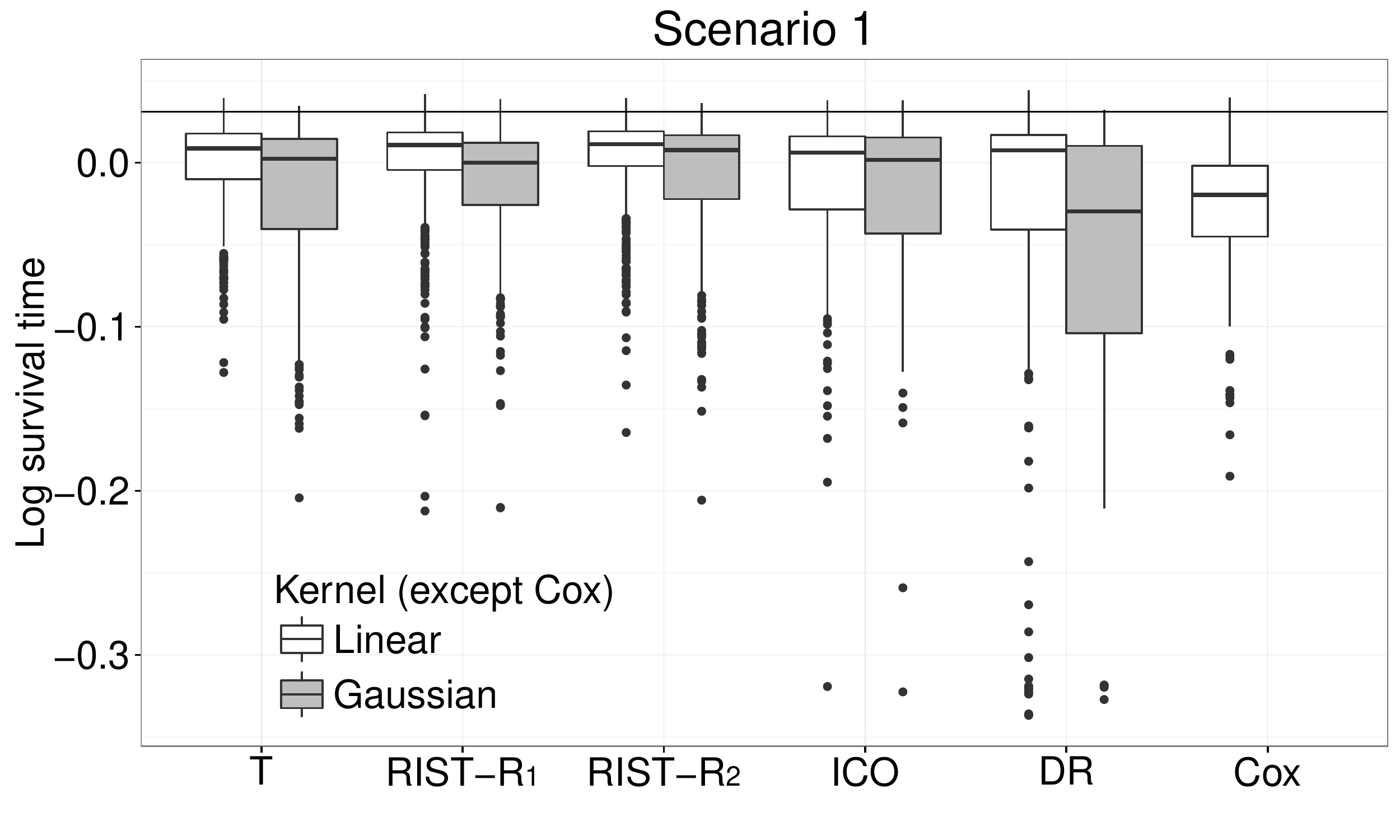} \\
   \includegraphics[trim=0in 0in 0in 0in, height=1.6in]{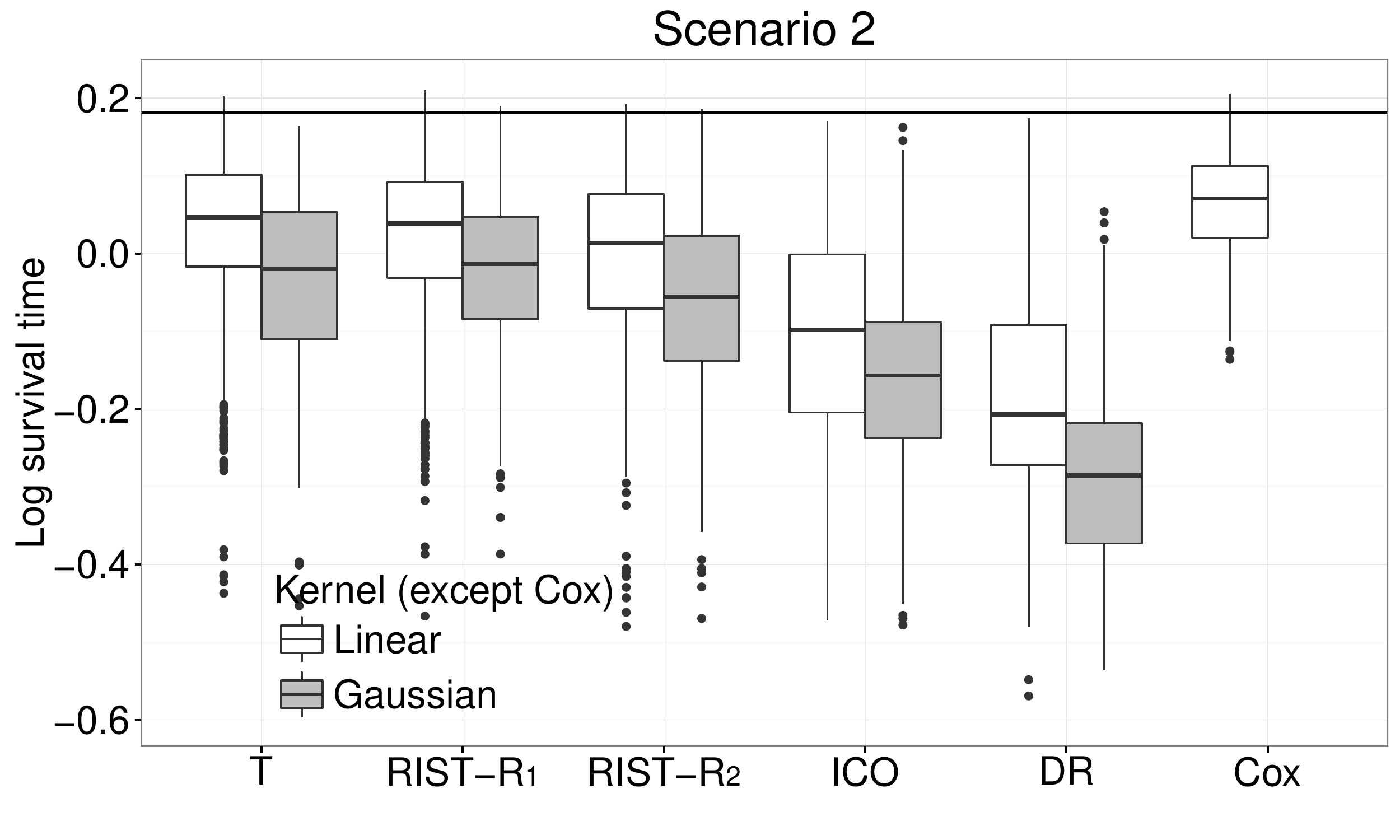} \\
   \includegraphics[trim=0in 0in 0in 0in, height=1.6in]{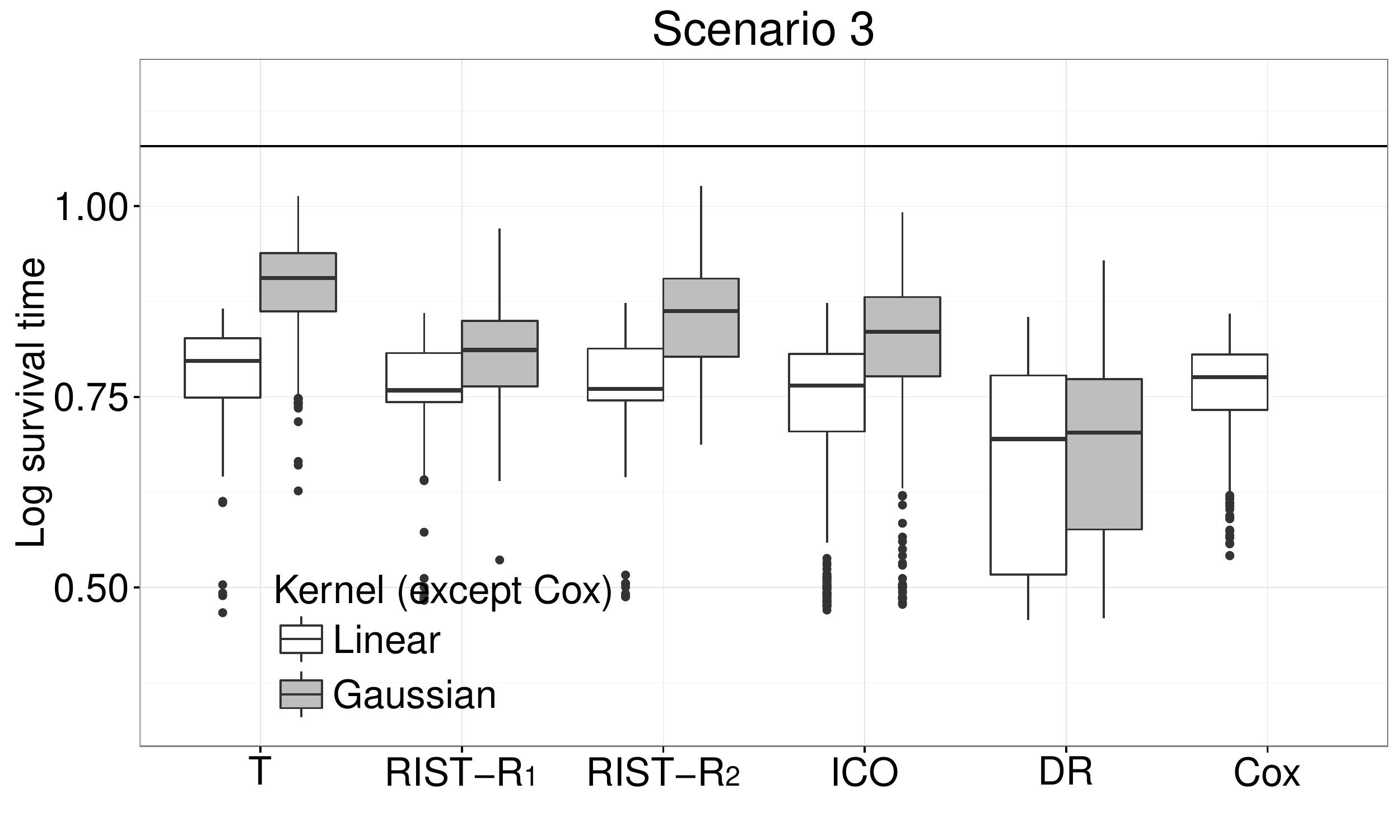} \\
   \includegraphics[trim=0in 0in 0in 0in, height=1.6in]{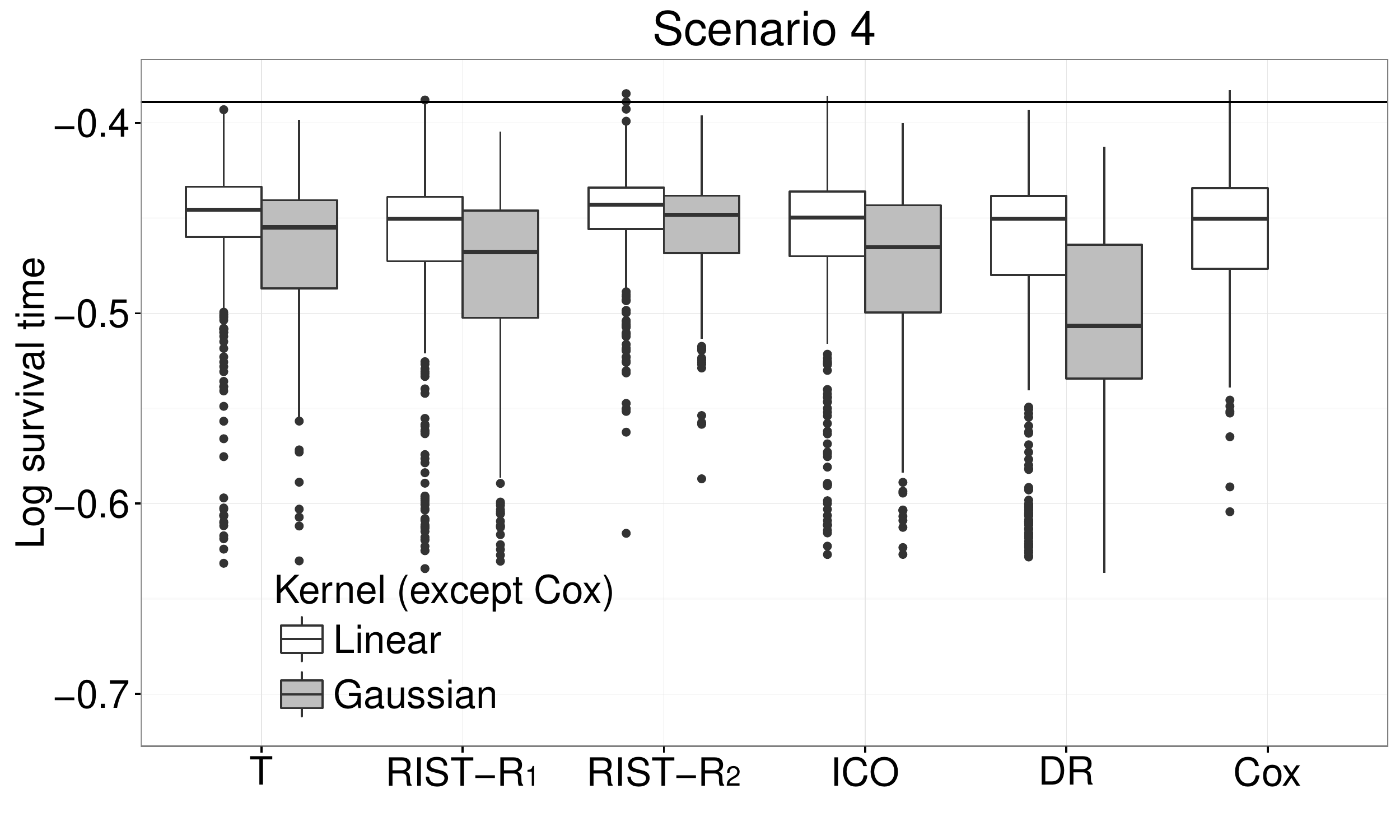} \\
    \caption{Boxplots of mean log survival time for different treatment regimes. Censoring rate: 30\%. T: using true survival time as weight; RIST-$R_1$ and RIST-$R_2$: using the estimated $R_1$ and $R_2$ respectively as weights, while the conditional expectations are estimated using recursively imputed survival trees;  ICO: inverse probability of censoring weighted learning; DR: doubly robust outcome weighted learning. The black horizontal line is the theoretical optimal value.}
    \label{fig3}
\end{figure}

\begin{figure}[H]
\centering
   \includegraphics[trim=0in 0in 0in 0in, height=1.6in]{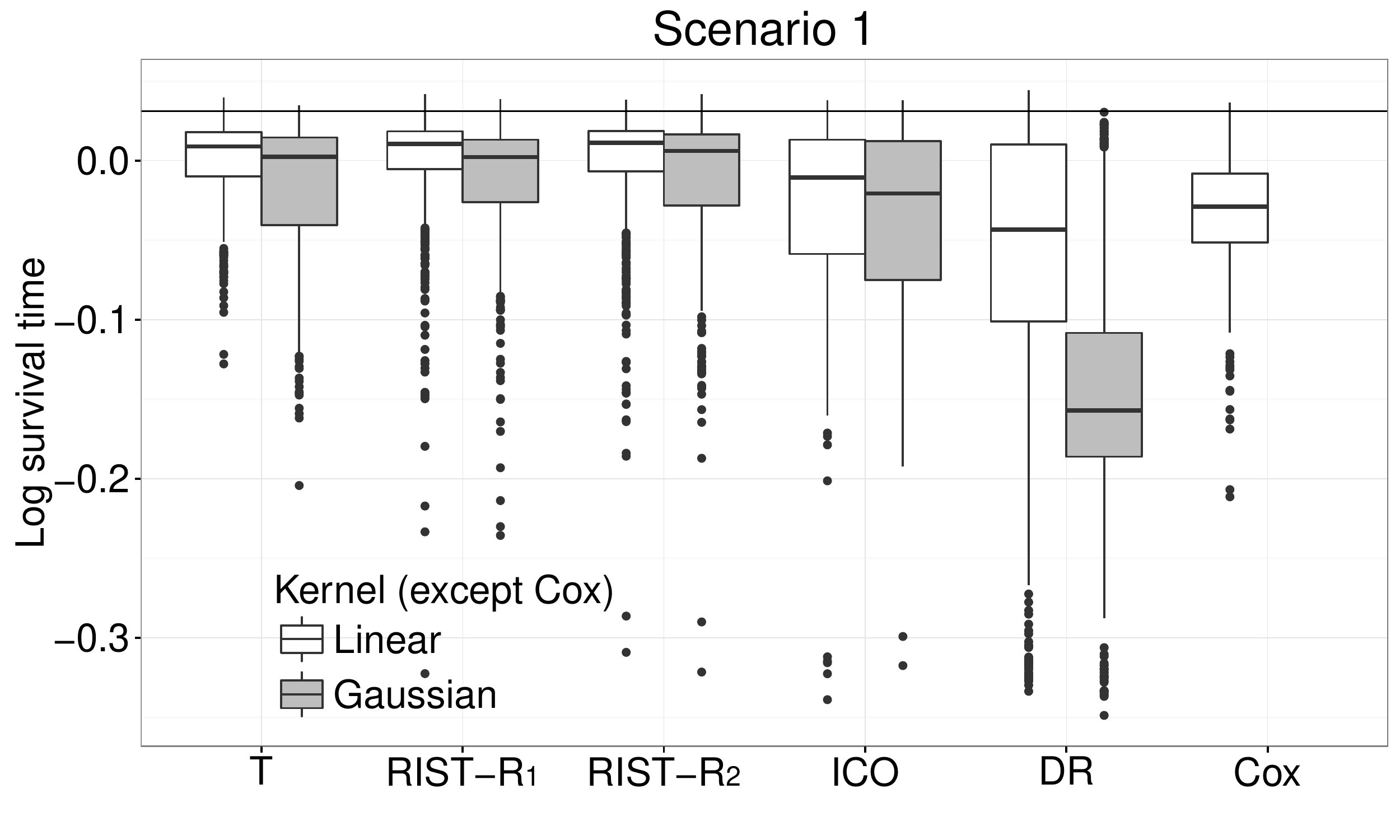} \\
   \includegraphics[trim=0in 0in 0in 0in, height=1.6in]{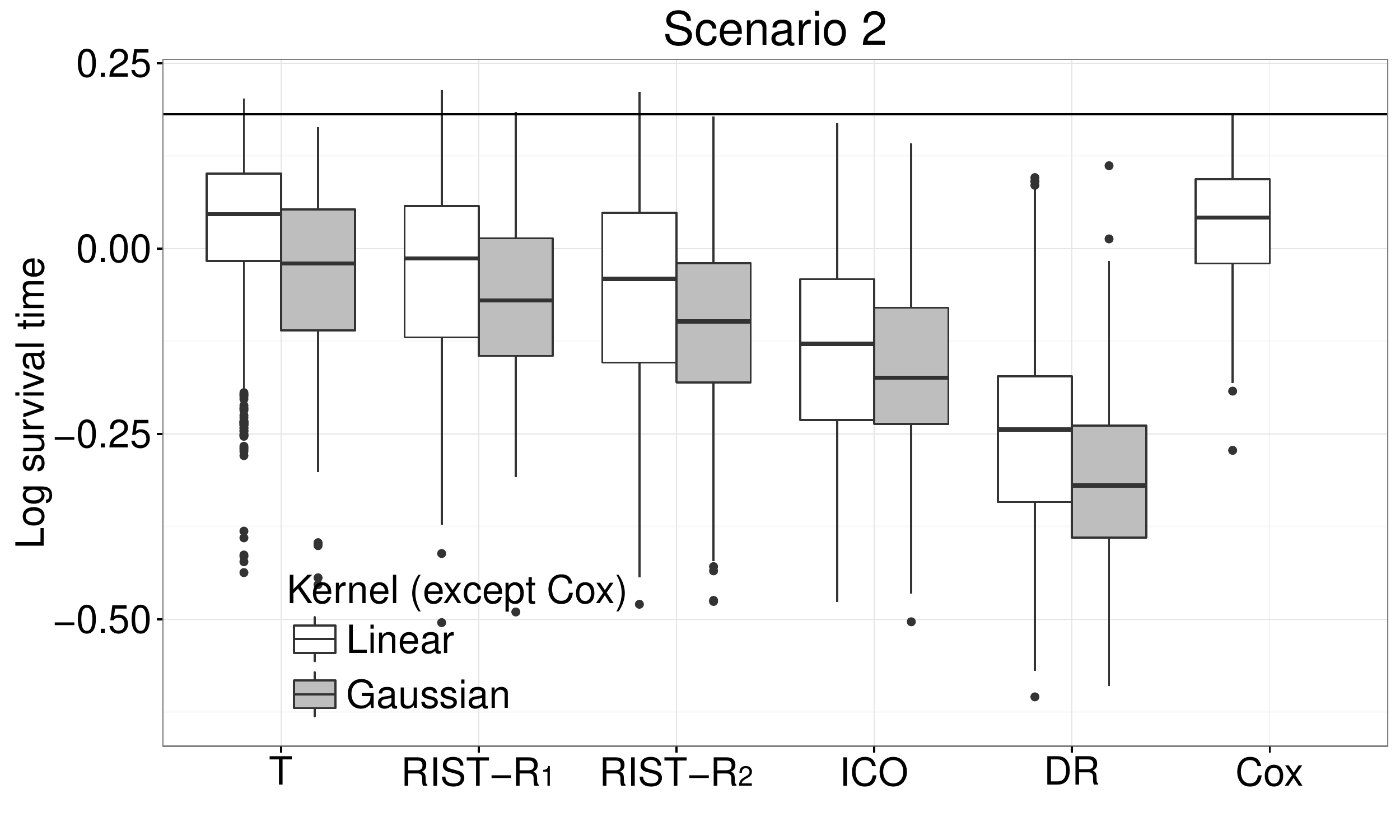} \\
   \includegraphics[trim=0in 0in 0in 0in, height=1.6in]{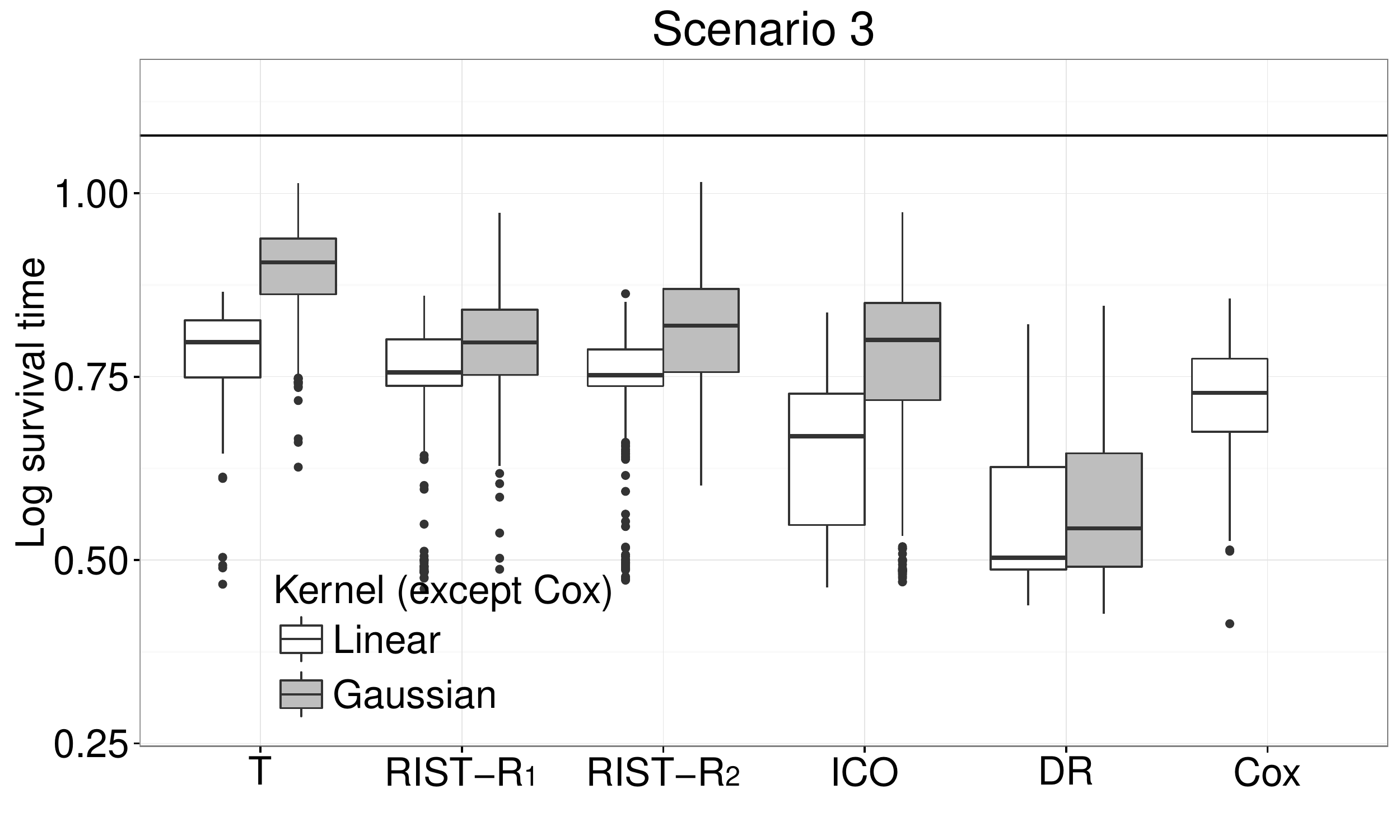} \\
   \includegraphics[trim=0in 0in 0in 0in, height=1.6in]{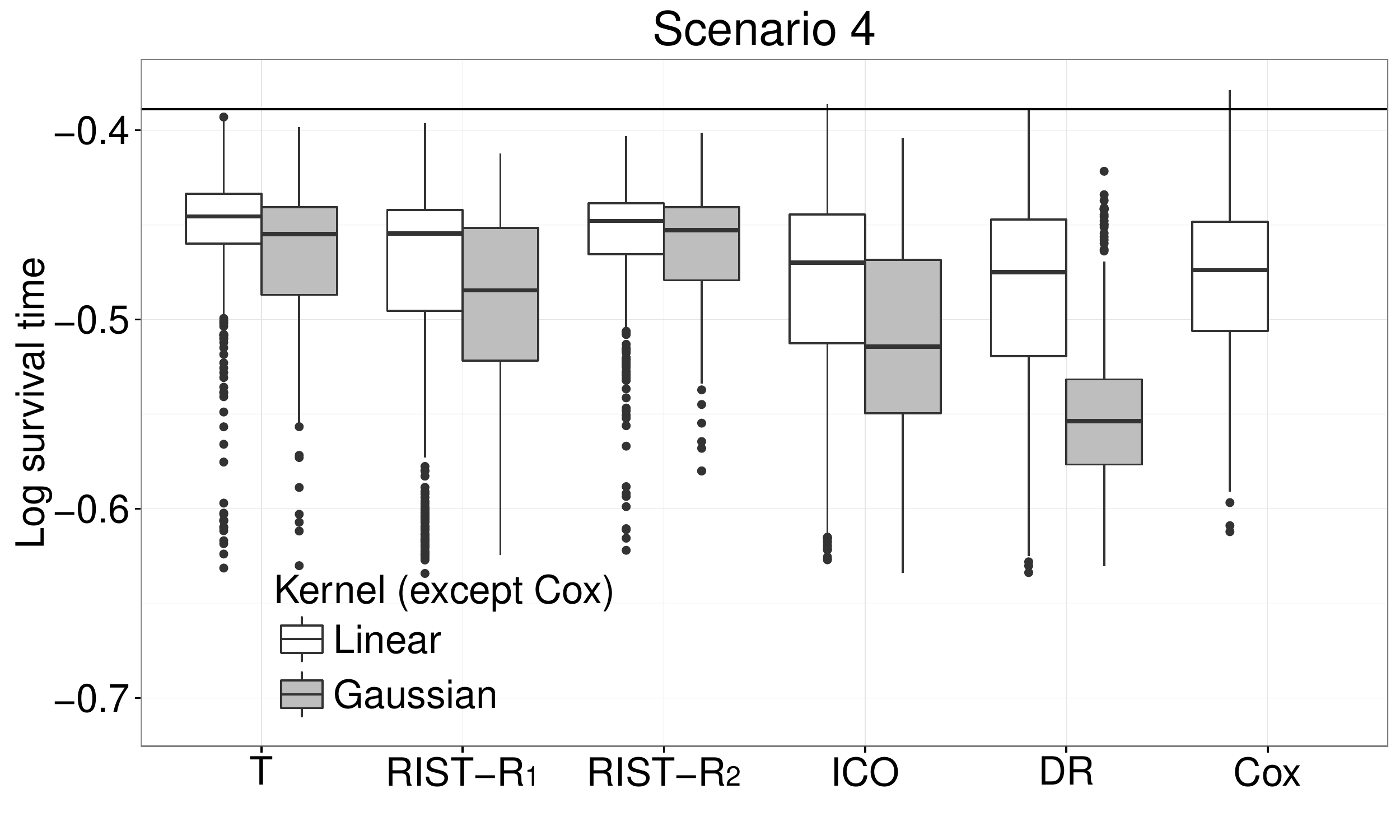} \\
    \caption{Boxplots of mean log survival time for different treatment regimes. Censoring rate: 60\%. T: using true survival time as weight; RIST-$R_1$ and RIST-$R_2$: using the estimated $R_1$ and $R_2$ respectively as weights, while the conditional expectations are estimated using recursively imputed survival trees;  ICO: inverse probability of censoring weighted learning; DR: doubly robust outcome weighted learning. The black horizontal line is the theoretical optimal value.}
    \label{fig4}
\end{figure}

We summarize the additional simulation results in this section. For each simulation scenario considered in Section \ref{section4}, we alter the first constant term in the censoring distribution to achieve 30\% (Table \ref{tab3} and Figure \ref{fig3}), and 60\% (Table \ref{tab4} and Figure \ref{fig4}) censoring rates.

\begin{table}[H]
\caption{\label{tab3}
Simulation results: Mean ($\times 10^3$) and (sd) ($\times 10^3$). Censoring rate: 30\%. For each scenario, the theoretical optimal value ($\times 10^3$) is 31, 181, 1079, and -389, respectively.}
\begin{tabular}{cccccccc}
\noalign{\smallskip}
\noalign{\smallskip}
 & kernel & T & RIST-$R_1$ & RIST-$R_2$ & ICO & DR & Cox\\
\noalign{\smallskip}
\multirow{2}{*}{1} & Linear & ~~~0 ~(26) & ~~~1 ~(31) & ~~~2 ~(28) & ~-10 ~(40) & ~-20 ~(63) & \multirow{2}{*}{~-26 (33)}\\
                & Gaussian  & ~-17 ~(44) & ~-10 ~(34) & ~~-7 ~(37) & ~-18 ~(45) & ~-48 ~(65)\\
\noalign{\smallskip}
\multirow{2}{*}{2} & Linear & ~~22 (113) & ~~17 (105) & ~-14 (126) & -110 (136) & -193 (133) & \multirow{2}{*}{~~65 (63)}\\
                & Gaussian  & ~-39 (115) & ~-25 (101) & ~-62 (113) & -164 (119) & -285 (112)\\
\noalign{\smallskip}
\multirow{2}{*}{3} & Linear & ~785 ~(52) & ~768 ~(53) & ~771 ~(52) & ~737 (95) & ~667 (124) & \multirow{2}{*}{~763 (61)}\\
                & Gaussian  & ~896 ~(61) & ~810 ~(54) & ~854 ~(69) & ~817 (124) & ~679 (123)\\
\noalign{\smallskip}
\multirow{2}{*}{4} & Linear & -453 ~(37) & -465 ~(46) & -448 ~(27) & -461 ~(42) & -471 ~(54) & \multirow{2}{*}{-457 (32)}\\
                & Gaussian  & -465 ~(35) & -477 ~(42) & -456 ~(27) & -474 ~(41) & -505 ~(48)\\
\noalign{\smallskip}
\end{tabular}
\source{T: using true survival time as weight; RIST-$R_1$ and RIST-$R_2$: using the estimated $R_1$ and $R_2$ respectively as weights, while the conditional expectations are estimated using recursively imputed survival trees; ICO: inverse probability of censoring weighted learning; DR: doubly robust outcome weighted learning; Cox: Cox proportional hazards model using covariate-treatment interactions.}
\end{table}

\begin{table}[H]
\caption{\label{tab4}
Simulation results: Mean ($\times 10^3$) and (sd) ($\times 10^3$). Censoring rate: 60\%. For each scenario, the theoretical optimal value ($\times 10^3$) is 31, 181, 1079, and -389, respectively.}
\begin{tabular}{cccccccc}
\noalign{\smallskip}
\noalign{\smallskip}
 & kernel & T & RIST-$R_1$ & RIST-$R_2$ & ICO & DR & Cox\\
\noalign{\smallskip}
\multirow{2}{*}{1} & Linear & ~~~0 ~(26) & ~~~-2 ~(39) & ~~~-5 ~(43) & ~-29 ~(57) & ~-64 ~(92) & \multirow{2}{*}{~-34 (36)}\\
                & Gaussian  & ~-17 ~(44) & ~-12 ~(40) & ~~-12 ~(45) & ~-35 ~(55) & ~-144 (78)\\
\noalign{\smallskip}
\multirow{2}{*}{2} & Linear & ~~22 (113) & ~-36 (123) & ~-61 (135) & -138 (133) & -248 (129) & \multirow{2}{*}{~~31 (79)}\\
                & Gaussian  & ~-39 (115) & ~-69 (108) & -102 (115) & -165 (117) & -313 (101)\\
\noalign{\smallskip}
\multirow{2}{*}{3} & Linear & ~785 ~(52) & ~753 ~(77) & ~748 ~(69) & ~646 (104) & ~556 (94) & \multirow{2}{*}{~721 (70)}\\
                & Gaussian  & ~896 ~(61) & ~796 ~(63) & ~819 ~(67) & ~775 (106) & ~573 (93)\\
\noalign{\smallskip}
\multirow{2}{*}{4} & Linear & -453 ~(37) & -478 ~(55) & -458 ~(33) & -486 ~(55) & -492 ~(59) & \multirow{2}{*}{-480 (43)}\\
                & Gaussian  & -465 ~(35) & -492 ~(48) & -461 ~(29) & -513 ~(53) & -551 ~(38)\\
\noalign{\smallskip}
\end{tabular}
\source{T: using true survival time as weight; RIST-$R_1$ and RIST-$R_2$: using the estimated $R_1$ and $R_2$ respectively as weights, while the conditional expectations are estimated using recursively imputed survival trees; ICO: inverse probability of censoring weighted learning; DR: doubly robust outcome weighted learning; Cox: Cox proportional hazards model using covariate-treatment interactions.}
\end{table}

\bibliographystyle{Chicago}
\bibliography{owlsurvival}

\end{document}